\documentclass[twocolumn,showpacs,preprintnumbers,amsmath,amssymb,eqsecnum,pre]{revtex4}

\usepackage{graphicx}      
\usepackage{dcolumn}       

\newcommand{\bm}{\mathbf}
\newcommand{\cm}{\mathcal}
\newcommand{\beq}{\begin{equation}}
\newcommand{\eeq}{\end{equation}}
\newcommand{\beqn}{\begin{eqnarray}}
\newcommand{\eeqn}{\end{eqnarray}}
\renewcommand{\Im}{\text{Im}}

\newcommand\beqa{\begin{eqnarray}}
\newcommand\eeqa{\end{eqnarray}}
\newcommand{\nn}{\nonumber\\}
\newcommand{\dd}{\text{d}}
\newcommand{\al}{\alpha}
\newcommand{\an}{\lambda}

\begin{document}
\title{Structure of hard-hypersphere fluids in odd dimensions}

\author{Ren\'e D. Rohrmann}
\email{rohr@oac.uncor.edu}
 \altaffiliation{ Permanent address:
Observatorio Astron\'omico, Universidad Nacional de C\'ordoba,
Laprida 854, X5000BGR C\'ordoba, Argentina}

\author{Andr\'es Santos}
\email{andres@unex.es} \homepage{http://www.unex.es/fisteor/andres}
\affiliation{ Departamento de F\'{\i}sica, Universidad de
Extremadura, E-06071 Badajoz, Spain}

\date{\today}
\begin{abstract}

The structural properties of single component fluids of hard
hyperspheres in odd space dimensionalities $d$ are studied with an
analytical approximation method that generalizes the Rational
Function Approximation  earlier introduced in the study of
hard-sphere fluids [S. B. Yuste and A. Santos, Phys.\ Rev.\ A {\bf
43}, 5418 (1991)]. The theory makes use of the exact form of the
radial distribution function to first order in density and extends
it to finite density by assuming a rational form for a function
defined in Laplace space, the coefficients being determined by
simple physical requirements. Fourier transform in terms of reverse
Bessel polynomials constitute the mathematical framework of this
approximation, from which an analytical expression for the static
structure factor is obtained. In its most elementary  form, the
method recovers the solution of the Percus--Yevick closure to the
Ornstein--Zernike equation for hyperspheres at  odd dimension. The
present formalism allows one to go beyond by yielding solutions with
thermodynamic consistency between the virial and compressibility
routes to any desired equation of state. Excellent agreement with
available computer simulation data at $d=5$ and $d=7$ is obtained.

\end{abstract}


\maketitle
\section{ Introduction} \label{s.intr}

Fluid systems made of hard bodies constitute simple  models in which
impenetrable particles interact solely through hard-core repulsions.
Despite their simplicity, hard-sphere systems are commonly used as
reference models to obtain accurate descriptions of real substances
(simple fluids, colloidal suspensions, granular media, and glasses)
over a wide range of state conditions \cite{hansen,barker,mulero}.
In particular, hard-sphere systems exhibit typical liquid-like
phenomena, such as a first-order freezing transition
\cite{alder,HR68,michels} and  metastable glass states \cite{S84}.

Hard-hypersphere fluids (where  the interaction potential is
infinite when two hyperspheres overlap and zero otherwise) are the
natural extension of hard spheres to arbitrary dimensions $d$. Such
systems have attracted an everlasting attention of many researchers
\cite{michels,freasier,LB82,J82,leutheusser,frisch,leutheusser2,rosenfeld87,wyler,baus,song,ASV89,luban,maeso,GGS91,LZKH91,frisch2,
Velasco,BMC99,parisi,YSH00,S00,GAL01,finken,robles,BMV04,CM04,L05,SH05,lue,bishop,BMV05,BW05,LB06,skoge,TS06,RHS07,BW07,
whitlock,scardicchio,P07}. The main reason is twofold. First,
studies of hard particles in high dimensions may reveal general
behaviors of the equation of state (EOS), radius of convergence of
the virial series, phase transitions, and fluid structure that can
help to understand the corresponding properties in real fluids.
Second, hypersphere systems provide well defined and very demanding
test models for theoretical approximations to many-body problems.

The thermodynamic and structural properties of $d$-sphere fluids in
high dimensions have been examined by computer simulations
\cite{michels,GAL01,robles,lue,bishop,BW05,LB06,skoge,BW07,whitlock}.
At a theoretical level, a number of virial coefficients have been
evaluated \cite{LB82,J82,BMV04,CM04,L05,BMV05}, the asymptotic
properties in the limit of infinitely many dimensions have been
investigated \cite{frisch,wyler,LZKH91,frisch2,parisi,SH05}, several
approximate EOS have been proposed
\cite{baus,song,ASV89,luban,maeso,GGS91,BMC99,YSH00,S00}, and
scaled-particle and density-functional methods  have been applied to
the fluid-solid phase transition \cite{Velasco,finken}. Regarding
the structural properties, the Percus--Yevick (PY) closure to the
Ornstein--Zernike (OZ) relation has been proven to be exactly
solvable at $d=\text{odd}$ \cite{freasier,leutheusser}, the solution
having been worked out at $d=5$ \cite{freasier} and $d=7$
\cite{robles,RHS07} {(apart from the classical cases of $d=1$
\cite{frenckel} and $d=3$ \cite{wertheim,thiele})}, and overlap
volume function representations have been proposed
\cite{leutheusser2,rosenfeld87}. Additionally, special interest has
focused on the packing problem and the formation of jammed
structures \cite{skoge,TS06,scardicchio,P07,torquato}.

The primary aim of this paper is to introduce an analytical method
for the study of hard-particle fluids in Euclidean spaces of
arbitrary odd dimension. Specifically, we derive an expression for a
key function directly related to the static structure factor $S(k)$
of the $d$-sphere fluid, from which all other structural and
thermodynamic properties can be expressed. Our technique is based on
the Rational Function Approximation (RFA) method, originally
developed for three-dimensional  hard spheres  \cite{yuste} and
applied to a wide variety of problems \cite{HYS07}, including
hard-sphere mixtures \cite{yuste_mixture}, sticky hard spheres
\cite{yuste_sticky,YS93,SYH98},  square-well fluids
\cite{yuste_square,AS01}, and penetrable spheres \cite{MYS07}. In
the RFA approach developed in this paper we define a Laplace-space
functional $G(s)$ of the radial distribution function $g(r)$ that
allows one to obtain   $S(k)$ in a simple way. By making use of the
exact form of $g(r)$ to first order in density, a function $\Psi(s)$
directly related to $G(s)$ is approximated  by  a Pad\'e
approximant, its coefficients being constrained by basic physical
conditions arising from the small wavenumber behavior of $S(k)$. The
simplest implementation of the approach, i.e., the one with an equal
number of coefficients and constraints, turns out to coincide with
the PY solution. The next extension contains two extra coefficients
that are fitted to reproduce any desired EOS in a thermodynamically
consistent way.

The paper is organized as follows. Section \ref{s.phys} summarizes
the basic physical tools involved in this study and describes the
application of reverse Bessel polynomials in the evaluation of
Fourier transforms in odd dimensional space.  An explicit expression
for the overlap volume between two hyperspheres, which plays a
prominent role in the generalization of the RFA method, is derived
in Sec.\ \ref{s.vol}. Section \ref{s.S} is devoted to the asymptotic
expressions of $S(k)$ and $G(s)$ for long wavenumber, for short
wavenumber, and for low densities. The generalization of the RFA
approach to arbitrary odd dimension is presented in Sec.\
\ref{s.RFA} and the corresponding evaluation of the direct
correlation function $c(r)$
 is given in Sec.\ \ref{s.cr}. Section
\ref{s.results} shows explicit results for fluids in dimensions
$d\le 11$ and comparisons with available computer simulations for
$d=5$ and $d=7$. The paper is closed with some concluding remarks in
Sec.\ \ref{s.concl}.

\section{Framework} \label{s.phys}

\subsection{Definitions}

The structure of a fluid is typically studied in terms of the radial
distribution function, $g(r)$, and a closely related function, the
structure factor $S(k)$, given by
\beq  \label{S}
S(k)=1+\rho \widehat{h}(k).
\eeq
Here $k$ is the wavenumber, $\rho$ is the density, and
$\widehat{h}(k)$ is the Fourier transform of the total correlation
function,
\beq \label{hg}
h(r)=g(r) -1.
\eeq
While $g(r)$ gives the relative probability of finding a particle
located a distance $r$ from another particle located at the origin,
$S(k)$ is proportional to the scattered intensity of radiation from
the fluid and thus is obtainable from scattering experiments. {An
additional useful quantity to describe  the fluid structure is the
direct correlation function $c(r)$, which is defined through the OZ
relation. In  Fourier space it reads }
\beq \label{ck}
\widehat{c}(k)=\frac{\widehat{h}(k)}{1+\rho
\widehat{h}(k)}=\frac{1}{\rho}\left[1-\frac{1}{S(k)}\right].
\eeq

The thermodynamics of hard $d$-sphere fluids can be fully accounted
for by the compressibility factor $Z\equiv p/\rho k_B T$
(dimensionless combination of  pressure $p$, density $\rho$, and
temperature $T$), which can be evaluated in turn from the contact
value of the radial distribution function, $g(\sigma^+)$,
\beq \label{Zgc}
Z = 1 + 2^{d-1} \eta  g(\sigma^+),
\eeq
where $\eta$ is the fraction of the total volume occupied by the
$d$-spheres (or packing fraction) and $\sigma$ the diameter of a
particle. The general relation between $\rho$ and $\eta$ reads
\beq \label{eta}
\eta=v_d \rho \sigma^d, \quad
v_d=\frac{(\pi/4)^{d/2}}{\Gamma(1+d/2)},
\eeq
where $v_d$ is the volume of a $d$-dimensional sphere of unit
diameter. In odd dimensions,
\beq \label{vd}
v_d =\frac{ (\pi/2)^{(d-1)/2}} {d!!}.
\eeq
Furthermore, the structure factor is related to thermodynamics
through the isothermal susceptibility as follows,
\beq \label{chi_S}
\chi \equiv k_B T \left(\frac{\partial \rho}{\partial p}\right)_T
=S(0).
\eeq
The relation between the isothermal susceptibility and the
compressibility factor is given by
\beq \label{chiZ}
\chi^{-1}=\frac{\dd}{\dd\eta} (\eta Z).
\eeq
This equation can be used  to impose thermodynamic consistency
between the virial and compressibility routes to the EOS.

For hard-particle systems in equilibrium, the structural properties
considered here are athermal, i.e., the temperature does not play
any relevant role. Moreover, the thermodynamic state can be
characterized by a variable alone, i.e., the density, the pressure,
or any related variable (e.g., $\eta$, $Z$, $\chi$).

\subsection{Low density expansions} \label{s.hard}

We consider briefly some exact low density results which will be
included in the analytical theory presented in this paper. We start
from the following general relationship
\beq
g(r)=\left[1+f(r)\right] y(r),
\eeq
where $y(r)$ is the cavity function and  $f(r)$ is the Mayer
function, which in the case of hard $d$-spheres reads
\beq \label{fo}
f(r)=\left\{ \begin{array}{rr} -1, & r< \sigma, \\ 0, & r> \sigma.\\
\end{array}\right.
\eeq
The low density behavior of $g(r)$ can be derived from the virial
expansion of the cavity function,
\beq
y(r)=1+y_1(r)\rho +y_2(r)\rho^2+\cdots,
\label{yvirial}
\eeq
where the functions $y_n(r)$ are represented by cluster diagrams
\cite{hansen,barker}. In particular, the first-order contribution to
the cavity function is
\beq \label{y1_r}
y_1(r) = \int \dd\bm{r}'\,f(r') f(|\bm{r}-\bm{r}'|) .
\eeq
It is worth noting that, because of Eq.\ (\ref{fo}), $y_1(r)$
represents the intersection volumen of two identical $d$-dimensional
spheres of radius  $\sigma$ whose centers are separated by a
distance $r$ \cite{noteB}.

With the preceding expressions, the lowest order terms in the
$\eta$-expansion of the radial distribution function,
\beq \label{g_eta}
g(r)=g_0(r)+g_1(r)\eta + O(\eta^2),
\eeq
are
\beq \label{g0_r}
g_0(r)= 1 + f(r) = \Theta(r-\sigma),
\eeq
\beq \label{g1_r}
g_1(r) = \frac{\sigma^{-d}}{v_d} \left[1 + f(r)\right] y_1(r) ,
\eeq
where $\Theta(x)$ is the step function [$\Theta(x)=1$ if $x>1$ and
zero otherwise]. The factor $1+f(r)$ prevents any pair of particle
centers from getting closer than a distance $\sigma$.

Fourier transformation of Eq.\ \eqref{g1_r} gives
\beq \label{g1_y}
\widehat{g}_1(k) = \frac{\sigma^{-d}}{v_d} \left[ \widehat{y}_1(k) +
\frac{1}{(2\pi)^d} \int \dd\bm{k}'\,\widehat{y}_1(k')
\widehat{f}(|\bm{k}-\bm{k}'|)  \right],
\eeq
where, by application of the convolution theorem in (\ref{y1_r}),
\beq \label{y1_k}
\widehat{y}_1(k) = \left[ \widehat{f}(k) \right]^2.
\eeq
 \subsection{Fourier transform in odd dimensions}

The Fourier transform of an absolutely integrable function
$\xi(\bm{r})$ in $d$ dimensions is defined by
\beq \label{a}
\widehat{\xi}(\bm{k})= \int \dd\bm{r} \,\xi(\bm{r}) e^{-i\bm{k\cdot
r}}
\eeq
and the associated inverse operation is given by
\beq \label{b}
\xi(\bm{r})= \frac{1}{(2\pi)^d} \int \dd\bm{k}
\,\widehat{\xi}(\bm{k}) e^{i\bm{k\cdot r}},
\eeq
where $\bm{k}$ is the wave vector. It is proven in Appendix
\ref{s.mathe} that, if the function $\xi(\bm{r})=\xi(r)$ depends
only on the magnitude $r=|\bm{r}|$ of the vector $\bm{r}$ and
$d=\text{odd}$, then the $d$-dimensional direct and inverse Fourier
transforms (\ref{a}) and \eqref{b} can be expressed as
\beq \label{fou_k}
\widehat{\xi}(k)= 2\frac{(2\pi)^{(d-1)/2}}{k^{d-2}} \Im
\left\{\mathcal{F}_n [\xi(r)] (-ik)\right\},
\eeq
\beq \label{fou_r}
\xi(r) =2 \frac{(2\pi)^{-(d+1)/2}}{r^{d-2}} \Im\left\{\mathcal{F}_n
            [\widehat{\xi}(k)] (-ir)\right\},
\eeq
respectively. Here, $\Im(z)$ denotes the imaginary part of $z$, $n$
is an integer related to $d$ by
\beq \label{n}
n\equiv \frac{d-3}2, \quad  d=2n+3,
\eeq
and   $\cm{F}_n[\xi(x)](s)$ is a functional of $\xi(x)$ defined by
\beq \label{Fs}
\cm{F}_n[\xi(x)](s) \equiv \int_0^\infty \dd x x \xi(x)
\theta_n(sx)e^{-sx} ,
\eeq
where the function $\theta_n(t)$ is the so-called {\em reverse
Bessel polynomial} of degree $n$
\cite{bochner,krall,carlitz,grosswald,weisstein}, whose expression
is
\beq \label{theta}
\theta_n(t)= \sum_{j=0}^{n} \omega_{n,j} t^j,\quad \omega_{n,j}=
\frac{(2n-j)!}{2^{n-j}(n-j)!j!}.
\eeq
Some of the properties of $\theta_n(t)$ are summarized in Appendix
\ref{s.mathe}. Table \ref{ta.1} provides the polynomials
$\theta_n(t)$ of degree less than seven.
\begin{table}
\caption{\label{ta.1} Reverse Bessel polynomials  of degree less
than seven.}
\begin{ruledtabular}
\begin{tabular}{cl}
$n$& $\theta_n(t)$\\
\hline
$0$&$1$ \\
$1$&$1+t$ \\
$2$&$3+3t+t^2$ \\
$3$&$15+15t+6t^2+t^3$ \\
$4$&$105+105t+45t^2+10t^3+t^4$ \\
$5$&$945+945t+420t^2+105t^3+15t^4+t^5$ \\
$6$&$10395+10395t+4725t^2+1260t^3+210t^4+21t^5+t^6$ \\
\end{tabular}
\end{ruledtabular}
\end{table}

Two useful applications  of Eq.\ \eqref{Fs} correspond to $\xi(x)=1$
and to the step function $\xi(x)=\Theta(x-1)$. With the help of
(\ref{rec2a}) and (\ref{theta}) one finds
\beq \label{F1}
\cm{F}_n[1](s) =\frac{\theta_{n+1}(0)}{s^2}=\frac{(2n+1)!!}{s^2},
\eeq
\beq \label{Ftheta}
\cm{F}_n[\Theta(x-1)](s) =\frac{\theta_{n+1}(s) e^{-s}}{s^2}.
\eeq

Henceforth we will indistinctly use $d$ and $n=(d-3)/2$ in the
remainder of the paper. Except for a few exceptions, we will
generally follow the rule of employing $n$  in subscripts and $d$ in
exponents.

\section{The two-hypersphere overlap volume} \label{s.vol}

{ The intersection volume of two $d$-spheres whose centers are
separated a distance $r$ is a key quantity in the study of hard
systems \cite{TS06,rosenfeld85}. Apart from its geometrical
interest, it yields the cavity function to first order in density,
$y_1(r)$, as mentioned below Eq.\ \eqref{y1_r}. Some expressions of
$y_1(r)$ in terms of special functions and recurrence relations can
be found in the literature \cite{rosenfeld87,baus,torquato03}, and
an explicit expression has been recently derived by Torquato and
Stillinger \cite{TS06}. In this Section we use the representation
\eqref{fou_r} for $d=2n+3=\text{odd}$ to provide an alternative
analytical expression of $y_1(r)$.}

 Taking into account that
$f(r)=\Theta(r-1)-1$, where henceforth we take $\sigma$ as the
length unit (i.e., $\sigma=1$)  and making use of Eqs.\
\eqref{fou_k},  \eqref{F1},  and \eqref{Ftheta}, one has
\cite{note1}
\beq \label{fSH_k}
\widehat{f}(k) = \frac{(2\pi)^{(d-1)/2}}{k^d}i\left[
\theta_{n+1}(-ik) e^{ik} -\theta_{n+1}(ik) e^{-ik}\right].
\eeq
{}From the property (\ref{limit}) it is easy to prove that, as
expected
\beq \label{y1_lim}
\lim_{k\rightarrow 0}\widehat{f}(k) = -\frac{2(2\pi)^{(d-1)/2}}{d!!}
=-2^d v_d.
\eeq

The contribution $y_1(r)$ to the cavity function is given by
\beq \label{y1_fou}
y_1(r) = \frac{(2\pi)^{-(d+1)/2}}{r^{d-2}} i
\int_{-\infty}^{\infty}\dd k\, k \widehat{y}_1(k) \theta_{n}(ikr)
e^{-ikr} ,
\eeq
where, according to (\ref{y1_k}) and (\ref{fSH_k}),
\beqn \label{y1_kth}
\widehat{y}_1(k) &=& \frac{(2\pi)^{d-1}}{k^{2d}}\left[ 2
\theta_{n+1}(ik)\theta_{n+1}(-ik) \right. \nn && \left. -
\theta_{n+1}^2(ik) e^{-2ik} - \theta_{n+1}^2(-ik) e^{2ik} \right].
\eeqn
Since $\widehat{y}_1(0)=(2^d v_d)^2=\mbox{finite}$, the integrand in
(\ref{y1_fou}) is regular along the integration interval and so we
can consider an integration path in the complex plane from
$k=-\infty$ to $k=+\infty$ that goes round the point $k=0$ from
below. The integral in Eq.\ (\ref{y1_fou}) decomposes into three
contributions with integrands headed by $e^{-ikr}$, $e^{-ik(r+2)}$
and $e^{-ik(r-2)}$, respectively. If $r>2$, we can close the path
with a lower half-circle of infinite radius, so that the three
contributions vanish. If $0<r<2$, however, the path in the third
integral must be closed with an upper half-circle and the residue
theorem yields a nonzero value. Therefore, for positive $r$ one has
\beqn \label{y1_fou3}
y_1(r)= \frac{(2\pi)^{(d-1)/2}}{r^{d-2}} R_{4n+4}(r) \Theta(2 -r),
\eeqn
where the residue
\beq \label{residue1}
R_{4n+4}(r)=\mathop{\mbox{Res}}_{k=0}\left[  k^{-(2d-1)}
\theta_{n+1}^2(-ik) \theta_n(ikr) e^{ik(2-r)}\right]
\eeq
 is given by the term of order $2d-2=4n+4$
in the $k$-expansion of $\theta_{n+1}^2(-ik)
\theta_n(ikr)e^{ik(2-r)}$, i.e.
\beq \label{residue2}
R_{4n+4}(r)=\left. \frac{1}{(4n+4)!} {\partial_t}^{4n+4}
\theta_{n+1}^2(-t) \theta_n(rt) e^{(2-r)t}\right|_{t=0},
\eeq
where we have made the change $t=ik$. Equation (\ref{residue2})
implies that $R_{4n+4}(r)$ is a polynomial of degree $4n+4$, as
indicated by the notation. Since $\theta_{n+1}^2(-t)\theta_n(rt)$ is
a polynomial of degree $3n+2$ in $t$, the Taylor expansion of the
exponential factor contributes to $R_{4n+4}(r)$ with factors
$(2-r)^j$ with $j\ge n+2$. Therefore, $R_{4n+4}(r)$ factorizes into
$(2-r)^{n+2}$ times a polynomial of degree $3n+2$. According to Eq.\
(\ref{y1_fou3}), the latter polynomial starts with
$r^{d-2}=r^{2n+1}$ because $y_1(r)$ must remain finite when
$r\rightarrow 0$. {}From this analysis, we obtain
\beq \label{residue}
R_{4n+4}(r)= r^{2n+1} (2-r)^{n+2} P_{n+1}(r),
\eeq
where $P_{n+1}(r)$ is a polynomial of degree $n+1$. An explicit
expression of $P_{n+1}(r)$ is given in Appendix \ref{s.overlap}. We
finally obtain the first-order contribution to the cavity function
for hard-sphere fluids in odd dimensions or, equivalently, the
overlap volume of two hyperspheres of {radius $\sigma=1$ with
centers separated a distance $r$ \cite{noteB}},
\beq \label{y1}
y_1(r) = \Theta(2 -r) (2\pi)^{(d-1)/2} (2-r)^{(d+1)/2} P_{n+1}(r).
\eeq
Notice that $y_1(0)$ is equivalent to the volume of {\em one}
hypersphere the radius $\sigma=1$, i.e., $y_1(0)= 2^d v_d$.
Moreover, setting $r=1$ in Eq.\ \eqref{Omega1} yields
\beq
\frac{y_1(1)}{y_1(0)}=1-\frac{(2n+3)!!}{2^{n+2}}\sum_{j=0}^{n+1}\frac{(-4)^{-j}}{
(2j+1)j!(n+1-j)!}.
\label{Omega(1)}
\eeq
This result  provides a simple expression for the third virial
coefficient $B_3$ of hard $d$-sphere systems in odd dimensions.
{}From the virial expansion
\beq \label{virial}
Z=1+\sum_{l=2}^\infty B_l \rho^{l-1}
\eeq
and Eqs.\ \eqref{Zgc} and \eqref{yvirial} one has $B_2=2^{d-1} v_d$
and $B_3=2^{d-1} v_d y_1(1)$. Therefore,
\beq \label{B3coeff}
\frac{B_3}{B_2^2} = 2\frac{y_1(1)}{y_1(0)}.
\eeq
%

\section{ Asymptotic behaviors of the structure factor} \label{s.S}

In this Section we examine the  asymptotic long  wavenumber, short
wavenumber, and low density  behaviors of the structure factor and
of a closely associated function, $G(s)$, which will play a central
role in the next Sections. We define $G(s)$ as the functional
$\cm{F}_n$, introduced in (\ref{Fs}), of the  radial distribution
function,
\beq \label{Gs}
G(s) \equiv \cm{F}_n[g(r)](s)=\int_0^\infty \dd r \,r g(r)
\theta_n(sr)e^{-sr}.
\eeq
We have found that this is the optimal  generalization to any odd
dimension of the Laplace transform $G(s)=\cm{L}[r g(r)](s)$ used in
Ref.\ \cite{yuste} in the study of hard spheres ($d=3$, $n=0$). With
(\ref{hg}) and (\ref{F1}), we note that
\beq \label{GH}
G(s) =\frac{(d-2)!!}{s^2} +\int_0^\infty \dd r\,r h(r)
\theta_n(sr)e^{-sr}.
\eeq
Hence, using (\ref{S}) and (\ref{fou_k}), the structure factor is
given by
\beq \label{S_G}
S(k)=1+\rho \frac{(2\pi)^{(d-1)/2}}{k^{d-2}} i  \left[ G(ik)-G(-ik)
\right].
\eeq
%

\subsection{Long and short wavenumber limits}

We shall now derive general conditions that the functions $G(s)$ and
$S(k)$ must satisfy. First, one can easily obtain an asymptotic
expression of $G(s)$ for {\em large} $s$ by replacing
$rg(r)=\Theta(r-1)\left[g(1^+)+O(r-1)\right]$  and $\theta_n(sr)=s^n
r^n\left[1+O(s^{-1})\right]$ in (\ref{Gs}). The result is
\beq
G(s)=g(1^+)s^{(d-5)/2}  {e^{- s}}[1+O(s^{-1})],
\eeq
i.e.
\beq \label{Gsgde}
\lim_{s\rightarrow \infty} s^{(5-d)/2} e^{s} G(s) = g(1^+).
\eeq
It follows from Eqs.\ \eqref{S_G} and \eqref{Gsgde} that, at long
wavenumber, $S(k)$ adopts the form
\beq \label{Skbig}
S(k) \approx 1+ \frac{2^d d!!}{ k^{(d+1)/2}}\eta g(1^+) \sin\left[k
+ \frac{\pi}{4}(5-d)\right].
\eeq
Therefore,  the structure factor for long $k$ oscillates with an
amplitude proportional to the contact value $g(1^+)$, an envelope
decaying as $k^{-(d+1)/2}$, and a phase shift equal to $\pi(5-d)/4$.

 On the other hand, the asymptotic form of $G(s)$ for
{\em small} $s$ can be derived from (\ref{GH}) by using Eq.\
(\ref{theta}) and the Taylor expansion of $e^{-sr}$. The result is
\beq
\label{GsH}
G(s)=\frac{(d-2)!!}{s^2} +\sum_{j=0}^\infty \al_{n,j} H_{j+1} s^{j},
\eeq
where
\beq
H_j\equiv\int_0^\infty \dd r\, r^j h(r)
\eeq
is the $j$th moment of the total correlation function and the
numerical coefficients $\al_{n,j}$ are given by
\beq  \label{Aj}
\al_{n,j} = \sum_{l=0}^{\min(n,j)} \frac{(-1)^{j-l}}{(j-l)!}
\omega_{n,l}.
\eeq
One can verify that the first $n$ coefficients $\al_{n,j}$ with
$j=2m+1=\text{odd}$ vanish, i.e.,
\beq  \label{Ajzero}
\al_{n,2m+1} = 0, \quad  m=0,1,\ldots,n-1.
\eeq
Therefore,
\beqa
 \label{GsHbis}
G(s)&=&\frac{(d-2)!!}{s^2} +\sum_{m=0}^\infty \al_{n,2m} H_{2m+1}
s^{2m}\nn &&+\sum_{m=n}^\infty \al_{n,2m+1} H_{2m+2} s^{2m+1}.
\eeqa
The property \eqref{Ajzero} is essential to guarantee that $S(k)$
remains bounded at the limit of zero wavenumber and, consequently,
the isothermal susceptibility takes finite values. Thus, inserting
Eq.\ \eqref{GsHbis} into Eq.\ \eqref{S_G}, we get the Taylor
expansion of $S(k)$ as
\beq
S(k)= 1-{2^dd!!\eta} \sum_{m=n}^\infty (-1)^m  \al_{n,2m+1} H_{2m+2}
k^{2(m-n)}.
\eeq
Application of Eq.\ \eqref{chi_S} provides a direct relationship
between the $(d-1)$th moment of the total correlation function and
the isothermal susceptibility, namely
\beq \label{chi_H}
\chi= 1+ {2^dd\eta} H_{d-1},
\eeq
where use has been made of the property
\beq
\al_{n,2n+1}=\frac{(-1)^{n+1}}{(2n+1)!!}. \label{A2n1}
\eeq

\begin{table*}
\caption{\label{ta.2} Polynomials $Q_{3n+4}(s)$  associated with
dimensions $d=3$, 5, 7, and 9.}
\begin{ruledtabular}
\begin{tabular}{cl}
$n$&$Q_{3n+4}(s)$\\
\hline $0$&
$12 + 12s -6 s^2 - 2 s^3 + (5/2) s^4$ \\
$1$&
$-2160-2160s-360s^2+360s^3+30s^4-42s^5+3s^6+(53/8) s^7$ \\
$2$& $1512000+1512000s+453600s^2-50400s^3-47880s^4+2520s^5
+2940s^6-300s^7-(285/2) s^8$\\
&$+(789/16) s^9+(289/16) s^{10}$ \\
$3$&
 $-2667168000-2667168000 s - 952560000 s^2 - 63504000s^3
+46720800s^4
+8618400s^5 -1738800s^6  $\\
& $-378000s^7 +69930s^8+11130s^9 -3045s^{10}
-(585/128)s^{11}+(38615/128)s^{12}
+(6413/128)s^{13}$ \\
\end{tabular}
\end{ruledtabular}
\end{table*}

\subsection{Low density expansion}

The series expansion of the radial distribution function in terms of
the packing fraction $\eta$, Eq.\ (\ref{g_eta}), leads to a similar
expansion for the function $G(s)$,
\beq \label{Geta}
G(s)= G_0(s)+G_1(s)\eta+O(\eta^2),
\eeq
with $G_j(s)=\cm{F}_n[g_j(r)](s)$. The zeroth order term is derived
from Eqs.\ (\ref{g0_r})  and (\ref{Ftheta}),
\beq \label{G0}
G_0(s) =s^{-2}{\theta_{n+1}(s) e^{-s}}.
\eeq
Next, since, $g(r)=\Theta(r-1)y(r)$ and $y_1(r)$ vanishes for $r\geq
2$, one can write
\beq \label{G1_y1}
G_1(s)= \frac{1}{v_d}\int_1^2 \dd r\,r y_1(r) \theta_n(sr)e^{-sr}.
\eeq
It is proven in Appendix \ref{app_G1} that
\beq \label{G1}
G_1(s)=
\frac{\an}{s^{d-2}}G_0^2(s)+\frac{e^{-s}}{s^{d+2}}Q_{3n+4}(s),
\eeq
where
\beq \label{aa}
\an\equiv(-1)^{(d-1)/2} 2^{d-1} d!!
\eeq
and $Q_{3n+4}(s)$ is a polynomial of degree $3n+4=(3d-1)/2$.
Explicit expressions of the first few polynomials $Q_{3n+4}(s)$ are
given in Table \ref{ta.2}.

\section{ Rational Function Approximation} \label{s.RFA}
The results presented in the preceding Sections are exact. In this
Section, we propose the extension  to hyperspheres in arbitrary odd
dimensions of the RFA, originally introduced in the study of
three-dimensional hard-sphere systems \cite{yuste,YHS96,HYS07}. The
main steps in the RFA can be summarized as follows: (i) a functional
$G(s)$ of the radial distribution function $g(r)$ is defined by a
suitable Laplace transformation such that $G(s)$ is simply related
to the structure factor $S(k)$; (ii) using as a guide the low
density form of $G(s)$, an auxiliary function $\Psi(s)$ is
introduced; (iii) the unknown function $\Psi(s)$ is approximated by
a rational function (or Pad\'e approximant), the degree difference
between the numerator and denominator polynomials being dictated by
the exact large-$s$ behavior of $G(s)$; (iv) finally, the
coefficients of the rational form for $\Psi(s)$ are determined by
requiring consistency with the exact small-$s$ behavior of $G(s)$.

According to Eq.\ \eqref{S_G},  the first step described above is
accomplished by the functional $G(s)$ defined by Eq.\ \eqref{Gs}.
Its large-$s$ and small-$s$ behaviors are given by Eqs.\
\eqref{Gsgde} and \eqref{GsHbis}, respectively. In order to continue
with the step (ii), let us rewrite Eq.\ \eqref{Geta} as
\beq \label{Gpsieta}
G(s)=\frac{s^{d-2}
}{\Psi_0(s)e^{s}}+\frac{s^{d-2}}{\Psi_0^2(s)e^{2s}}
\left[\an-\Psi_1(s)e^s\right]\eta +O(\eta^2),
\eeq
where, according to Eqs.\ \eqref{G0} and \eqref{G1},
\beq \label{psi0}
\Psi_0(s) =\frac{s^d}{\theta_{n+1}(s)},
\eeq
\beq \label{psi1}
\Psi_1(s) = - \frac{Q_{3n+4}(s)}{\theta_{n+1}^2(s)^2}.
\eeq
This suggests the introduction of the auxiliary function $\Psi(s)$
through
\beq \label{Grfa}
G(s) = \frac{s^{d-2}}{\Psi(s) e^{s}-\an\eta},
\eeq
so that
\beq
\Psi(s)=\Psi_0(s)+\Psi_1(s)\eta+O(\eta^2).
\eeq
The large-$s$ and small-$s$ conditions \eqref{Gsgde} and
\eqref{GsHbis} imply that
\beq
\lim_{s\to\infty} s^{-(d+1)/2}\Psi(s)=\frac{1}{g(1^+)},
\label{Psi_slarge}
\eeq
\beq
\left.\partial_s^j\Psi(s)e^s\right|_{s=0}=\begin{cases}
\an\eta,&j=0,\\
0,&1\leq j\leq d-1,\\
d!/(d-2)!!,&j=d,\\
0,&j=d+1+2m, \\
&m=0,\ldots,n,\\
\end{cases}
\label{Psi_ssmall}
\eeq
respectively, where in Eq.\ \eqref{Psi_ssmall} we have taken into
account Eq.\ \eqref{Ajzero}.

Thus far, Eqs.\ \eqref{Gpsieta}--\eqref{Psi_ssmall} are exact. Now
we follow step (iii) and approximate the auxiliary function
$\Psi(s)$ by a rational form (or Pad\'e approximant),
\beq \label{pade}
\Psi(s) =\frac{B_N(s)}{A_M(s)}, \quad A_M(s)\equiv\sum_{j=0}^M a_j
s^j,\quad B_N(s)\equiv\sum_{j=0}^N b_j s^j.
\eeq
Note that the choice of a rational form for $\Psi(s)$ is compatible
with the exact requirements \eqref{psi0}, \eqref{psi1},
\eqref{Psi_slarge}, and \eqref{Psi_ssmall}.
The combination of Eqs.\ (\ref{Grfa}) and (\ref{pade}) constitute a
simple approximation of $G(s)$ that will be made consistent with the
basic physical requirements outlined in Section \ref{s.S}. To begin
with, we note that the condition (\ref{Psi_slarge}) fixes the
relation between the degrees of the polynomials $A_M(s)$ and
$B_N(s)$, namely
\beq \label{NMd}
N=M+\frac{d+1}2.
\eeq
In fact, the ratio of the highest coefficients $a_M$ and $b_N$
directly gives the contact value of the pair distribution function:
\beq \label{abg}
g(1^+)=\frac{a_M}{b_N}.
\eeq

To close the RFA proposal \eqref{pade} we need to determine the
coefficients $\{a_j\}$ and $\{b_j\}$. Since one of them  can be
arbitrarily chosen, the number of independent unknowns is $N+M+1$.
Following the step (iv), we resort to the exact small-$s$ behavior
\eqref{Psi_ssmall}, which imposes $d+n+2=3n+5$ constraints.
Therefore, in order to make the problem solvable, one must have
$N+M\geq d+n+1=3n+4$. In view of Eq.\ \eqref{NMd}, this implies that
\beq \label{NM}
N\ge d, \quad M\ge n+1=\frac{d-1}{2}.
\eeq
The constraints stemming from Eq.\ \eqref{Psi_ssmall} or,
equivalently, Eq.\ \eqref{GsHbis} are worked out in Appendix
\ref{s.Pade}.

\subsection{Low density expansion}

It is worthwhile noting that the lower bounds \eqref{NM} can also be
derived by requiring consistency of the RFA form \eqref{pade} with
the exact zeroth order term, Eq.\ \eqref{psi0}, in a density
expansion. In the approximation with the least number of unknowns,
i.e., with
\beq \label{NMup}
N= d, \quad M= n+1,
\eeq
Eq.\  \eqref{psi0} implies that
\beq  \label{aj_0}
\left.\frac{b_j}{b_d}\right|_{\eta=0} =\delta_{j,d},\quad
\left.\frac{a_j}{b_d}\right|_{\eta=0} = \omega_{n+1,j},
\eeq
where $\delta_{j,d}$ is Kronecker's delta. As mentioned before, one
of the coefficients in (\ref{pade}) can be given a constant,
arbitrary non-zero value. Of course, this choice will not have any
consequence on the physical results derived from (\ref{Grfa}), but
an appropriate one may simplify the algebra involved in the further
development below. In this sense, in view of \eqref{aj_0}, two
adequate alternative choices are either
\beq \label{a0}
a_0\equiv \omega_{n+1,0}=(d-2)!!
\eeq
or
\beq \label{bN}
b_d \equiv 1.
\eeq
Henceforth we will adopt (\ref{a0}). With the assumption
\eqref{NMup} and the choice (\ref{a0}), it is seen from Eq.\
(\ref{psi1}) that the derivatives $a_j'\equiv\left.\partial_\eta
a_j\right|_{\eta=0}$ and $b_j'\equiv\left.\partial_\eta
b_j\right|_{\eta=0}$  obey the following equations:
\beq  \label{bj_der}
b_j' = -\frac{1}{(d-2)!!} \left[ q_{3n+4,j}
+\sum_{m=\max(0,j-M)}^{j-1} \omega_{n+1,j-m}  b_{m}' \right],
\eeq
for $0\le j\le N=d$, and
\beq \label{aj_der}
a_j' = q_{3n+4,N+j} +\sum_{m=j}^{M} \omega_{n+1,m} b_{N+j-m}',
\eeq
for $1\le j \le M=n+1$, where  $q_{3n+4,j}$ is the $j$th coefficient
of the polynomial $Q_{3n+4}(s)$. Application of (\ref{bj_der})
allows one to obtain recursively all the quantities $b_j'$, which
can  then be used in the evaluation of $a_j'$ with the help of
(\ref{aj_der}).

In conclusion, the analytical form provided by Eqs.\ (\ref{Grfa})
and (\ref{pade}) exactly reproduces the zeroth- and first-order
terms in density of $G(s)$ when suitable values of the Pad\'e
coefficients and their derivatives are used. In the next subsection,
we carry out the extension of this representation to arbitrary
densities.

\subsection{Standard approximation} \label{s.finite}

Let us consider the RFA form \eqref{pade} when the number of
unknowns ($N+M+1$) equals the number of constraints ($3n+5$). Taking
into account Eq.\ \eqref{NMd}, this corresponds to the choice
\eqref{NMup}. We will refer to this case as  the {\em standard} RFA,
i.e., the one in which all the Pad\'e coefficients are determined
from the basic constraints and so no free parameters remain. As will
be shown in Sec.\ \ref{s.cr}, this standard RFA turns out to provide
the exact solution of the PY integral equation for odd dimensions
\cite{freasier,leutheusser} by a completely different method.

For the case (\ref{NMup}), Eq.\ (\ref{b_new}) expresses the
coefficients $\{b_j\}$ in terms of the coefficients $\{a_j\}$.
Insertion  into Eqs.\ (\ref{Grfa}) and (\ref{pade}) allows one to
rewrite $G(s)$  in the form
\beq \label{Grfa_1}
G(s)= \frac{e^{- s}}{s^2} \frac{A_{n+1}(s) } { 1 + \an \eta
\sum_{j=0}^{n+1} a_j\phi_{d-j}(s)  } ,
\eeq
where we have called
\beq  \label{phi}
\phi_m(s) \equiv \frac 1{s^{m}} \left[ \sum_{j=0}^m
\frac{(-s)^j}{j!} -e^{-s} \right].
\eeq
Note that $\phi_m(0)=0$. Equation \eqref{Grfa_1} assumes the
normalization choice \eqref{a0} \cite{s.bN1}.

In order to evaluate the $n+1$ coefficients $\{a_j,
j=1,\ldots,n+1\}$ we can use the remaining constraints obtained from
(\ref{eqppal}). An equivalent procedure results from the
substitution of (\ref{Grfa_1}) into (\ref{GsH}), followed by a new
power-series analysis. This latter method is simpler than the other
one because the coefficients $\{b_j\}$ have already been eliminated.
In this case, the analysis of powers yields
\beq \label{DD}
D_l + \sum_{j=0}^{l-2} \gamma_j  D_{l-2-j} = \widetilde{a}_l,
\eeq
where we have called
\beq \label{agamma}
\widetilde{a}_j\equiv \frac{a_j}{(d-2)!!}, \quad \gamma_j \equiv
\frac{\al_{n,j}H_{j+1}}{(d-2)!!}.
\eeq
The coefficients $D_l$ are linear combinations of the $\{a_j\}$
given by $D_0=1$ and
\beq \label{Dj}
D_l= \frac{1}{l!} -\an\eta \sum_{m=0}^{n+1} t_{d-m,l} a_m, \quad
l\ge 1,
\eeq
where
\beq  \label{tnj}
t_{m,j}= \sum_{l=1}^j \frac{(-1)^{l+m}}{(l+m)!  (j-l)!}
\eeq
is the $j$th coefficient in the power series expansion of
$-e^s\phi_m(s)$. In Eq.\ \eqref{DD} and in the remainder of this
Section we have adopted the conventions $a_j= 0$ if $j>n+1$ and
$\sum_{j=0}^m \cdots=0$ if $m<0$.

Because of (\ref{Ajzero}), we have
\beq  \label{gammanull}
\gamma_1=\gamma_3=\cdots=\gamma_{2n-1}=0.
\eeq
Therefore, Eq.\ (\ref{DD}) with  $l=\text{even}\leq 2n+2$ can be
used to express the quantities $\gamma_{2m}$ with $0\le m \le n$ in
terms of the coefficients $a_j$ by means of the recursion relation
\beq  \label{eg0}
\gamma_{2m}= \widetilde{a}_{2m+2} -D_{2m+2}
-\sum_{j=0}^{m-1}\gamma_{2j} D_{2(m-j)},\quad 0\le m \le n.
\eeq
The parameters $\gamma_0$, $\gamma_2$, $\gamma_4$, $\gamma_6$,
\ldots are linear, quadratic, cubic, quartic, \ldots in the
coefficients $\{a_j\}$.
 Next, Eq.\ (\ref{DD}) with  $l=\text{odd}\leq 2n+1$ yields
\beq  \label{ec1}
D_{2m+1}+\sum_{j=0}^{m-1} \gamma_{2j} D_{2(m-j)-1} =
\widetilde{a}_{2m+1}, \quad 0\leq m\leq n.
\eeq
When the $\{\gamma_{2m}\}$ obtained from Eq.\ \eqref{eg0} are
inserted into Eq.\ \eqref{ec1} one gets a closed set of $n+1$
equations for $a_1$, $a_2$, \ldots, $a_{n+1}$. Therefore, the
implementation of the standard RFA method reduces to solving a set
of $n+1=(d-1)/2$ algebraic equations, which become nonlinear for
$n\geq 1$ or, equivalently, $d\ge 5$. In general, the number of
mathematical solutions (including complex ones) is $2^n$. In the
case of multiple solutions, we choose the solution
 which yields the correct low density limit given by
(\ref{aj_0}).
%
\begin{figure}
{\includegraphics[width=.9\columnwidth]{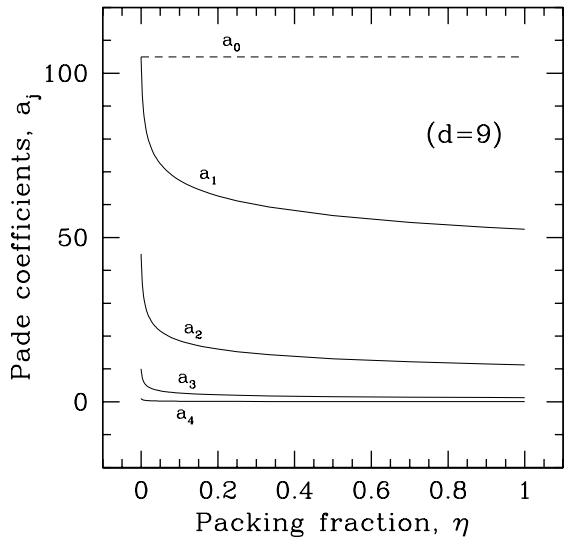}}
\caption{\label{f.am_d} Coefficients $a_j$, as a function of $\eta$,
obtained from the standard RFA method  for a fluid at $d=9$.}
\end{figure}
%
In this sense, it is important to note that the asymptotic behaviors
of $a_j$ and $b_j$ as $\eta \rightarrow 0$ [Eqs.\ (\ref{aj_0}),
(\ref{bj_der}), and (\ref{aj_der})] are naturally included among the
solutions of Eqs. (\ref{ec1}). This is due to two reasons: (a) Eq.\
(\ref{GsHbis}) is verified by the exact density expansion of $G(s)$
up to any order, in particular to zeroth and first orders [Eq.\
(\ref{Geta})];  (b)  the number of constraints derived from
(\ref{GsHbis}) coincides with the least number of Pad\'e
coefficients
 required in the exact expansion of $G(s)$ up to first
order in density.

The set of equations \eqref{ec1} can be analytically solved for
$d=1$, 3, 5, and 7, the solutions for $d=1$, 3, and 5 being
explicitly given in Appendix \ref{s.solutions}.  For
$d=\text{odd}\geq 9$, however, the solutions must be obtained
numerically. By means of explicit evaluations  in all the cases
analyzed, we have found that the coefficients $\{a_j\}$ are finite
for $0\leq \eta\leq 1$ and adopt the following values at $\eta=1$:
\beq  \label{aeta1}
\left.a_j\right|_{\eta=1} =2^{-j} \left.a_j\right|_{\eta=0}=
\frac{\omega_{n+1,j}}{2^j}.
\eeq
Figure \ref{f.am_d} illustrates the physical roots of Eq.\
(\ref{ec1}) for a fluid of dimension $d=9$ ($n=3$) as functions of
the packing fraction. It may be observed that, at the low density
limit ($\eta \rightarrow 0$), each $a_j$ tends to the corresponding
coefficient $\omega_{4,j}$ of the reverse Bessel polynomial
$\theta_4(t)$ (see Table \ref{ta.1}), as dictated by Eq.\
(\ref{aj_0}). These roots [except $a_0$ which is fixed by Eq.\
(\ref{a0})] monotonically decrease as the density increases,
reaching their minimum values given by Eq.\ (\ref{aeta1}) at
$\eta=1$.

Once  the coefficients $\{a_j\}$ are determined as functions of
$\eta$, the structure factor of the system is given by Eqs.\
(\ref{S_G}) and (\ref{Grfa_1}).  The ansatz (\ref{pade}) thus
provides  an explicit expression for the structure factor to any
finite density. Because the approximation to $G(s)$ is exact to
first order in density, the structure factor thus obtained is exact
to second order in density. In conclusion, the $4n+5$ constraints
derived from the small-$s$ behavior of $G(s)$, together with the
zero-density conditions (\ref{aj_0}), allow us to completely
characterize $S(k)$ for a hypersphere fluid in odd dimension in the
standard RFA approach. Since $S(k)$ and $\widehat{h}(k)$ are
directly related through Eq.\ \eqref{S}, application of Eq.\
\eqref{fou_r} yields the radial distribution function $g(r)$.

The compressibility factor resulting from the virial EOS (\ref{Zgc})
is determined by the standard RFA method through the relations
(\ref{abg})  and (\ref{b_new}). The result is
\beq  \label{zvirial}
Z_{v}(\eta)=1+ 2^{d-1}\eta a_{n+1}
\left[1+\an\eta\sum_{j=0}^{n+1}\frac{(-1)^{d-j}}{(d-j)!} a_j
\right]^{-1}.
\eeq
The isothermal susceptibility $\chi$ given by (\ref{chi_H}) can also
be easily evaluated. On the one hand, from  Eq.\ (\ref{A2n1}) and
the definition of the $\gamma_j$ factors [Eq.\ (\ref{agamma})] one
has
\beq
 \label{gamm_2}
H_{d-1}  = {(-1)^{n+1} }{[(d-2)!!]^2}\gamma_{d-2} .
\eeq
On the other hand, Eq.\ (\ref{DD}) at $l=d$ yields
\beq \label{gamm_1}
\gamma_{d-2}= -\Bigl( D_d +\sum_{j=0}^{n}\gamma_{2j}
D_{d-2-2j}\Bigr).
\eeq
Therefore, from Eq.\ (\ref{chi_H}) we obtain
\beq  \label{chiRFA}
\chi(\eta)=1-2\lambda(d-2)!!\eta \Bigl( D_d
+\sum_{j=0}^{n}\gamma_{2j} D_{d-2-2j}\Bigr),
\eeq
with factors $\gamma_0, \gamma_2,\ldots, \gamma_{d-3}$ given by
(\ref{eg0}). Once $\chi$ is known, the thermodynamic relation
(\ref{chiZ}) can be integrated to obtain $Z$ in the so-called
compressibility route,
\beq  \label{zcom}
Z_{c}(\eta)=\int_0^1 {\dd x}{\chi^{-1}(\eta x)}.
\eeq
Additionally, one finds from (\ref{aeta1}) the following limit
values
\beq  \label{zveta1}
Z_{v}(1)= \infty , \quad \chi(1)=0.
\eeq
Comparison between the results obtained from Eqs.\ (\ref{zvirial})
and (\ref{zcom}), which gives a measure of the degree of
thermodynamic inconsistency in the standard RFA solution, will be
presented in Sec.\ \ref{s.results}.

\subsection{Extended approximation} \label{improved}

It is possible to construct   RFA solutions more elaborate than the
standard RFA one by considering in Eq.\ \eqref{pade}  a number of
unknowns $N+M+1$ larger than the number $3n+4$ of basic constraints
\eqref{Psi_ssmall}. We will refer to this case as the
\emph{extended} RFA. The simplest extension corresponds to
\beq
N=d+1,\quad M=n+2,
\label{NM_ext}
\eeq
since Eq.\ \eqref{NMd} must be preserved. This involves two new
parameters ($a_{n+2}$ and $b_{d+1}$) which can be freely chosen
without compromising the basic physical requirements. A natural
choice is to adjust $a_{n+2}$ and $b_{d+1}$ by requiring prescribed
values of the contact value of the radial distribution function,
$g_c\equiv g(1^+)$, and of the isothermal susceptibility $\chi$. In
practice, one can use (\ref{chiZ}) to evaluate $\chi$ from $g_c$ (or
vice versa), so that only one EOS is needed and the thermodynamic
consistency between the virial and the compressibility routes is
thus guaranteed by construction.

With the choice \eqref{NM_ext}, the relationship (\ref{abg}) can be
used to eliminate one of the two new Pad\'e coefficients (for
instance, $b_{d+1}$) in terms of the other one ($a_{n+2}$). Making
use again of Eq.\ \eqref{b_new}, the {extended} RFA for $G(s)$ reads
\beq \label{Grfa_2}
G(s)= \frac{e^{- s}}{s^2}  \frac{A_{n+2}(s)}
{1+g_c^{-1}a_{n+2}s+\an\eta\sum_{j=0}^{n+2}a_j \phi_{d-j}(s) }.
\eeq
The set of Pad\'e coefficients $\{a_j,0\le j\le n+2\}$, is obtained
in a manner similar to that given above. Thus, $a_0$ remains fixed
by (\ref{a0}), while $a_1$, \ldots, $a_{n+1}$ are related to
$a_{n+2}$ by the $n+1$ equations (\ref{ec1}). So far, by setting
$a_{n+2}=0$ we recover the standard RFA. However, now we fix the
prescribed $\chi$ and then Eq.\ (\ref{chiRFA}) provides the needed
equation to close the set (\ref{ec1}).

The nonlinearity of the problem in the extended RFA is higher than
in the standard case, the number of mathematical solutions being
$2^{n+1}$. In particular, one has to deal with a quadratic equation
for $d=3$ \cite{yuste,YHS96,HYS07}, a quartic equation for $d=5$,
and so on. It can be verified that in the zero-density limit the
physical solution has the form
\beq
\left. a_j\right|_{\eta=0}=\omega_{n+1,j}+\left.
a_{n+2}\right|_{\eta=0}\omega_{n+1,j-1},
\label{aj0}
\eeq
\beq
\left. b_j\right|_{\eta=0}=\delta_{j,d}+\left.
a_{n+2}\right|_{\eta=0}\delta_{j,d+1},
\label{bj0}
\eeq
where the numerical value of $\left. a_{n+2}\right|_{\eta=0}$
depends on the value of the fourth virial coefficient predicted by
the prescribed EOS. Equations \eqref{aj0} and \eqref{bj0} imply that
\beq
\lim_{\eta\to 0}A_{n+2}(s)=\theta_{n+1}(s)\left(1+\left.
a_{n+2}\right|_{\eta=0}s\right),
\eeq
\beq
\lim_{\eta\to 0}B_{d+1}(s)=s^d\left(1+\left.
a_{n+2}\right|_{\eta=0}s\right),
\eeq
so that Eq.\ \eqref{psi0} is recovered,  irrespective of the chosen
EOS (provided, of course, it is consistent with the exact second and
third virial coefficients).

It is important to note  that the physical root of the set of
equations \eqref{ec1} and \eqref{chiRFA} must correspond to
$a_{n+2}>0$. Otherwise, since $a_0=(d-2)!!>0$, there would exist at
least one positive real root $s_0$ of the polynomial $A_{n+2}(s)$.
According to Eq.\ \eqref{Grfa_2}, this would imply $G(s_0)=0$, what
is incompatible with a positive definite $g(r)$. A careful analysis
of Eqs.\ \eqref{ec1} and \eqref{chiRFA} shows that the condition
$a_{n+2}>0$ requires that the chosen values of $Z$ and $\chi$
satisfy the inequalities $Z>Z_v^{\text{PY}}$ and
$\chi>\chi_c^{\text{PY}}$, where $Z_v^{\text{PY}}$ and
$\chi_c^{\text{PY}}$ are the compressibility factor and the
isothermal susceptibility, respectively, obtained from the standard
RFA or, equivalently, from the PY solution. Therefore, if the
prescribed $\chi$ is obtained from the prescribed $Z$ by application
of Eq.\ \eqref{chiZ}, the extended RFA provides  physical
correlation functions only if
\beq
Z_v^{\text{PY}}<Z<Z_c^{\text{PY}}.
\label{Zchi}
\eeq
When this condition is verified, the physical root of the set of
equations \eqref{ec1} and \eqref{chiRFA}  corresponds to the
smallest positive real value of $a_{n+2}$.

\section{ Direct correlation function} \label{s.cr}

According to Eq.\ \eqref{fourr}, the direct correlation function
$c(r)$ can be evaluated from its Fourier transform as
\beq \label{cr_fou}
c(r) = \frac{(2\pi)^{-(d+1)/2}}{r^{d-2}} i
\int_{-\infty}^{\infty}\dd k\, k \widehat{c}(k) \theta_{n}(ikr)
e^{-ikr} .
\eeq
In the RFA approach, using Eqs.\ \eqref{S_G}, (\ref{Grfa}), and
\eqref{pade}, we obtain
\beq  \label{SkAB}
S(k)
=\frac{B^{+}B^{-}-A^{+}A^{-}}{(B^{+}e^{ik}-A^{+})(B^{-}e^{-ik}-A^{-})},
\eeq
where $A^{\pm}\equiv\an\eta A_M(\pm ik)$ and $B^{\pm}\equiv B_N(\pm
ik)$. Next, use of Eq.\ (\ref{ck}) yields
\beq  \label{ckAB}
\rho \widehat{c}(k) =\frac{A^{+}B^{-}e^{-ik}+A^{-}B^{+}
e^{ik}-2A^{+}A^{-}}{B^{+}B^{-}-A^{+}A^{-}}.
\eeq
Since, according to Eqs.\ \eqref{GsH} and \eqref{Grfa}, one has
$B_N(s)e^s-\an\eta A_M(s)=s^d\left[1+O(s)\right]$, it follows that
the denominator in Eq.\ \eqref{SkAB} is of order $k^{2d}$.
Therefore, the numerator must also be of order $k^{2d}$ to have a
finite value of $S(0)$. More specifically,  for both the standard
($N=d$, $M=n+1$) and the extended  ($N=d+1$, $M=n+2$) RFA
approaches, one has
\beq  \label{BBAA2}
B^{+}B^{-}-A^{+}A^{-}= (b_{d-1}b_{d+1}-b_d^2) (ik)^{2d} +b_{d+1}^2
(ik)^{2d+2}.
\eeq

\subsection{Standard RFA}
In the case of the standard RFA, one has $b_{d+1}=0$, so that
$B^{+}B^{-}-A^{+}A^{-}= -b_d^2(ik)^{2d}$. Consequently, when
inserting Eq.\ \eqref{ckAB} into Eq.\ \eqref{cr_fou}, each one of
the three integrands  has  a pole of order $2d-1$ at $k=0$. A
residue analysis similar to that employed in Sec.\ \ref{s.vol} leads
to
\beq \label{cr_PY}
c(r) = \frac{ (-1)^{(d+1)/2} }{r^{d-2}}
{\overline{R}_{4n+4}(r)}\Theta(1-r),
\eeq
where $\overline{R}_{4n+4}(r)$ is a polynomial of degree $4n+4=2d-2$
given by
\beq \label{Rr_}
\overline{R}_{4n+4}(r)=\mathop{\mbox{Res}}_{s=0}\left[ \frac
{A_{n+1}(-s)B_{d}(s)} {b_d^2 s^{2d-1}} \theta_{n}(sr) e^{s(1-r)}
\right].
\eeq
Comparison with Eq.\ \eqref{y1_fou3} shows that, in the standard
RFA, $c(r)$ has in the region $0\leq r<1$ a polynomial form similar
to that of $y_1(r)$ in the region $0\leq r\leq 2$. As happened in
the latter case, finiteness of $c(0)$ implies that $c(r)$ is a
polynomial of degree $4n+4-(d-2)=d$.

Equation \eqref{cr_PY} shows that $c(r)=0$ for $r>1$ in the standard
RFA. Of course, the standard RFA complies with the physical
requirement $g(r)=0$ for $r<1$. These two conditions define
precisely the PY closure to solve the OZ equation for hard
hyperspheres. Therefore, we find that the standard RFA, i.e., the
approximation given by Eqs.\ (\ref{Grfa}) and (\ref{pade}) with the
least number of coefficients satisfying the requirements
(\ref{Gsgde}) and (\ref{GsHbis}), coincides with the PY solution for
hard hyperspheres of odd dimensions \cite{freasier,leutheusser}.
This is a remarkable result since both approaches are in principle
rather independent. In fact, following the philosophy behind the RFA
approach, it is straightforward to proceed to the  first natural
extension of the PY solution or standard RFA.

\subsection{Extended RFA}

In the extended RFA approach, $b_{d+1}=g_c^{-1} a_{n+2}\ne 0$. As a
consequence, Eq.\ \eqref{BBAA2} shows that, besides the pole of order
$2d-1$ at $k=0$, each one of the three contributions to
$k\widehat{c}(k)$ has also two simple poles $k=\pm i \kappa$,
$\kappa$ being a real quantity given by
\beq \label{kappa}
\kappa\equiv \sqrt{ (b_d^2-b_{d-1}b_{d+1}) b_{d+1}^{-2} }.
\eeq
Therefore, applying the residue theorem we get
\beq \label{cr_ex}
 c(r) = {(-1)^{(d+1)/2}}  (K_{+} + K_{-}-K_0)
\frac{\theta_n(\kappa r)e^{-\kappa r}} {r^{d-2}}
\eeq
for $r>1$ and
\beqa \label{cr_in}
 c(r) &=& -\frac{(-1)^{(d+1)/2} } { r^{d-2}}
\left[{\widetilde{R}}_{4n+4}(r)+ K_{-} \theta_n(-\kappa r)e^{\kappa
r} \right. \nn && \left. +(K_{0}-K_-) \theta_n(\kappa r)e^{-\kappa
r} \right]
\eeqa
for $r<1$. In the above expressions,
\beq \label{Kpm}
 K_{\pm}\equiv \frac{A_{n+2}(\mp \kappa)B_{d+1}(\pm \kappa)e^{\pm \kappa }}
 {2b_{d+1}^2\kappa^{2d}},
\eeq
\beq \label{K_a}
 K_0 \equiv\an\eta \frac{A_{n+2}(\kappa)A_{n+2}(-\kappa)}{b_{d+1}^2\kappa^{2d}},
\eeq
\beq \label{Rr_ex}
{\widetilde{R}}_{4n+4}(r) \equiv \mathop{\mbox{Res}}_{s=0}\left[
\frac {A_{n+2}(-s)B_{d+1}(s)} {b_{d+1}^2 s^{2d-1}(s^2-\kappa^2)}
\theta_{n}(sr) e^{s(1-r)} \right].
\eeq

Equation \eqref{cr_ex} shows that, in contrast to the PY solution
(or, equivalently, the standard RFA), $c(r)$ does not vanish outside
the core in the extended RFA. In fact, $c(r)\propto \theta_n(\kappa
r) e^{-\kappa r}/r^{d-2}$ for $r>1$. This functional form can be
considered as an extension to odd $d$ of the well-known Yukawa form
$e^{-\kappa r}/r$ for $d=3$ \cite{noteYuk}.

The form \eqref{cr_ex} obtained from the extended RFA for hard
hyperspheres admits an alternative interpretation. Imagine an
interaction potential with a hard core at $r=1$ plus an attractive
Yukawa tail $-\epsilon\theta_n(z r) e^{-z r}/r^{d-2}$ for $r>1$. In
the mean spherical approximation (MSA) \cite{hansen} the closure to
the OZ relation would be $c(r)=(\epsilon/k_BT)\theta_n(z r) e^{-z
r}/r^{d-2}$ for $r>1$, which has the same form as  Eq.\
\eqref{cr_ex}. As a consequence, the MSA for the hard-core Yukawa
interaction with $d=\text{odd}$ is exactly solvable, the
corresponding functions $G(s)$, $S(k)$, and $c(r)$ for $r<1$  being
given by Eqs.\ \eqref{Grfa_2}, \eqref{SkAB} (with $N=d+1$ and
$M=n+2$), and \eqref{cr_in}, respectively. The main difference
between the extended RFA for hard hyperspheres and the MSA for the
hard-core Yukawa potential is that, while in the former case the two
input parameters are the contact value $g_c=g(1^+)$ and the
isothermal susceptibility $\chi$, in the latter case the control
parameters are the reduced temperature $k_BT/\epsilon$ and the
inverse interaction range $z$.

In the three-dimensional case, the extended RFA \cite{yuste}
reproduces the the so-called generalized mean spherical
approximation (GMSA) for hard spheres \cite{waisman}, which consists
of closing the OZ equation with the assumption that $c(r)$  has a
Yukawa tail outside the core ($r>1$). {}From that point of view, the
extended RFA applied to higher odd dimensions can be seen as the
natural extension of the GMSA to those dimensions.

\section{ Results} \label{s.results}

\subsection{Standard RFA (PY theory)}

We list in Appendix \ref{s.solutions} the explicit solutions of Eq.\
(\ref{ec1}) corresponding to the standard RFA approach (or,
equivalently, the PY theory) for fluids in dimensions $d=1,3,5$. The
solution for $d=7$ is also analytical and agrees with the results
reported in Refs.\ \cite{robles,RHS07}.  The solutions to Eq.\
(\ref{ec1}) for $d=9, 11$ have been obtained numerically.

%
\begin{figure}
{\includegraphics[width=.9\columnwidth]{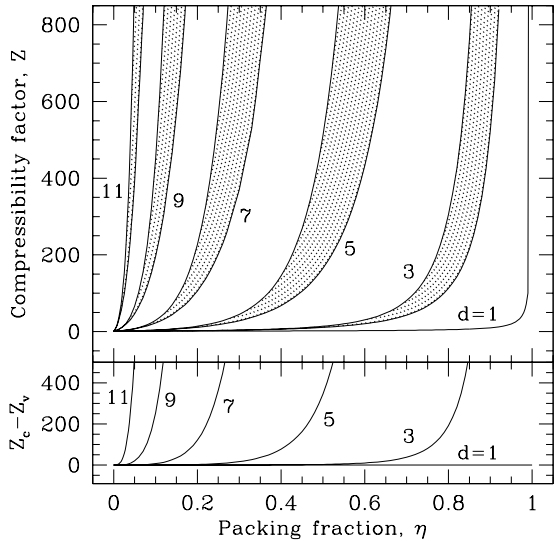}}
\caption{\label{f.zcompres} Top panel: Compressibility factors
$Z_v^{\text{PY}}$ and $Z_c^{\text{PY}}$ (curves on the right and
left edges of each shaded area, respectively) arising from the
virial and the compressibility routes according to the standard RFA
method  (PY theory)  at dimensions $d=1$, 3, 5, 7, 9, and 11, as
functions of the packing fraction $\eta$. Bottom panel: Difference
between $Z_c^{\text{PY}}$ and $Z_v^{\text{PY}}$. }
\end{figure}
%
\begin{table}
\caption{\label{ta.3} Packing fractions at freezing transition
($\eta_f$) and closest packing ($\eta_{c}$) predicted for fluids at
$d=1,3,5,7,9,11$.}
\begin{ruledtabular}
\begin{tabular}{ccc}
$d$ &  $\eta_f$ & $\eta_{c}$ \\
\hline
$1$ &  --    & $1$ \\
$3$ &  $0.494$\footnotemark[1]
    & $\sqrt{2}\pi/6\approx0.74048$\footnotemark[5] \\
$5$ & $0.19$\footnotemark[2]
    & $\sqrt{2}\pi^2/30\approx0.46526$\footnotemark[5] \\
$7$ & $0.072$\footnotemark[3]
    & $\pi^3/105\approx 0.29530$\footnotemark[5] \\
$9$ & $0.027$\footnotemark[4]
    & $\sqrt{2}\pi^4/945\approx0.14578$\footnotemark[5] \\
$11$& $0.009$\footnotemark[4]
    & $32\pi^5/(93555\sqrt{3})\approx 0.06043$\footnotemark[5] \\
\end{tabular}
\end{ruledtabular}
\footnotetext[1]{Monte Carlo simulations \protect\cite{HR68}}
\footnotetext[2]{Molecular dynamics simulations
\protect\cite{michels}} \footnotetext[3]{Molecular dynamics
simulations \protect\cite{robles}} \footnotetext[4]{Estimated by the
method of Refs.\ \protect\cite{finken,Velasco}}
\footnotetext[5]{Densest lattice packing listed in Ref.\
\protect\cite{web}.}
\end{table}
%

The compressibility factors $Z_v^{\text{PY}}$ and $Z_c^{\text{PY}}$
derived by the standard RFA approach from the virial and
compressibility  routes [Eqs. (\ref{zvirial}) and (\ref{zcom})],
respectively, are shown in Fig.\ \ref{f.zcompres} for
$d=1,3,5,7,9,11$, as functions of the packing fraction. Both routes
yield identical and exact results only in the case $d=1$.
Discrepancies between $Z_c^{\text{PY}}$ and $Z_v^{\text{PY}}$ grow
noticeably with increasing dimension  for $d\ge 3$ (bottom panel in
Fig.\ \ref{f.zcompres}). The compressibility factors predicted by
the standard RFA  have a singularity at $\eta=1$ for all $d$ [cf.\
Eq.\ (\ref{zveta1})]. However, since $d$-spheres are not space
filling (except for $d=1$) the true pressure must present a
singularity  at a certain density lower than or equal to the closest
packing fraction $\eta_c$. The maximal packing fractions presently
known in these dimensions are listed in the third column of Table
\ref{ta.3}.
%
\begin{figure}
{\includegraphics[width=.9\columnwidth]{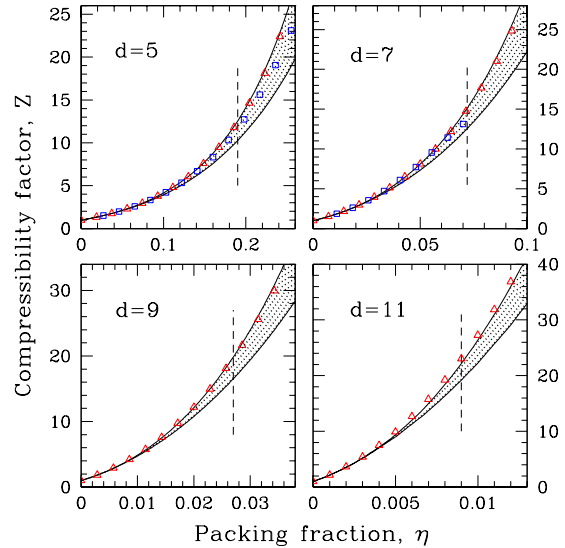}}
\caption{\label{f.z_dvarios} (Color online) Compressibility factors
$Z_c$ and $Z_v$ arising from the standard RFA (edges  of shaded
areas) for $d=5,7,9,11$.  The values obtained from the SMS EOS
\cite{song} for $d=5,7,9,11$ (triangles), and the LM EOS
\cite{luban} for $d=5,7$ (squares) are also shown. The vertical
dashed lines mark the freezing packing fraction reported in Table
\protect\ref{ta.3}. }
\end{figure}
%
In practice, the inconsistency between $Z_v^{\text{PY}}$ and
$Z_c^{\text{PY}}$ is not as severe as it is appears in Fig.\
\ref{f.zcompres} because the fluid phase is actually limited to very
low values of $\eta$ for high dimensions. Freezing transitions are
predicted for hard $d$-sphere fluids with $d>1$ and found to occur
at increasing lower packing fractions with increasing dimensionality
\cite{michels,finken,skoge}. Some values of the freezing packing
fraction $\eta_f$ are listed in the second column of Table
\ref{ta.3}. Figure \ref{f.z_dvarios} depicts the compressibility
factors $Z_c^{\text{PY}}$
 and $Z_v^{\text{PY}}$ up to densities in the neighborhood
of the liquid-solid phase transition  for fluids with $d=5,7,9,11$.
The figure also includes the values of $Z$ predicted by the
extension of the Carnahan--Starling EOS proposed by Song, Mason, and
Stratt (SMS) \cite{song}, as well as the predictions of the
semi-empirical EOS proposed by Luban and Michels (LM) \cite{luban}.
There is a reasonable agreement among these two latter EOS and the
PY results via the compressibility route within the fluid phase. In
fact, comparison of
 $Z_v^{\text{PY}}$ and $Z_c^{\text{PY}}$ with computer
simulations in the cases $d=5$ \cite{S00} and $d=7$ \cite{robles}
shows that, in the stable fluid region, the true values of $Z$ are
bracketed by $Z_v^{\text{PY}}$ and $Z_c^{\text{PY}}$ in the form
indicated by Eq.\ \eqref{Zchi}, being closer to $Z_c^{\text{PY}}$
than to $Z_v^{\text{PY}}$. This in passing gives further support to
the extended RFA, which requires as input an EOS satisfying Eq.\
\eqref{Zchi} in order to provide  physically meaningful results.

%
\begin{figure}
{\includegraphics[width=.9\columnwidth]{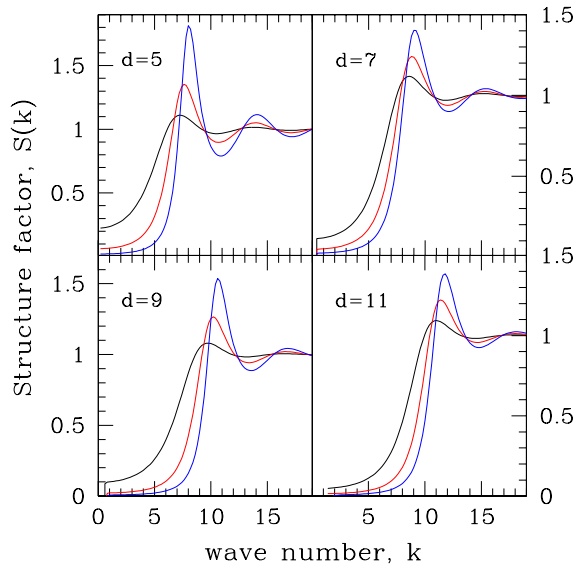}}
\caption{\label{f.Sk_PY_RL} (Color online) The structure factor as
obtained from the standard RFA (or PY solution) for hyperspheres in
$d=5$ (at $\eta=0.06,0.13,0.2$), $d=7$ (at $\eta=0.03,0.05,0.07$),
$d=9$ (at $\eta=0.01,0.025,0.04$), and $d=11$ (at
$\eta=0.005,0.01,0.015$). The first peak increases with increasing
density.}
\end{figure}
%
Figure \ref{f.Sk_PY_RL} shows the structure factor obtained from the
standard RFA method for fluids in dimensions $d=5,7,9,11$ for
densities near the values of $\eta_f$ given in Table \ref{ta.3}. As
expected, the oscillations of $S(k)$ become more pronounced as the
density increases. We can also observe that the location of the
first peak tends to move to higher wavenumbers as the dimensionality
increases \cite{skoge}.

\subsection{Extended RFA}


%
\begin{figure}
{\includegraphics[width=.9\columnwidth]{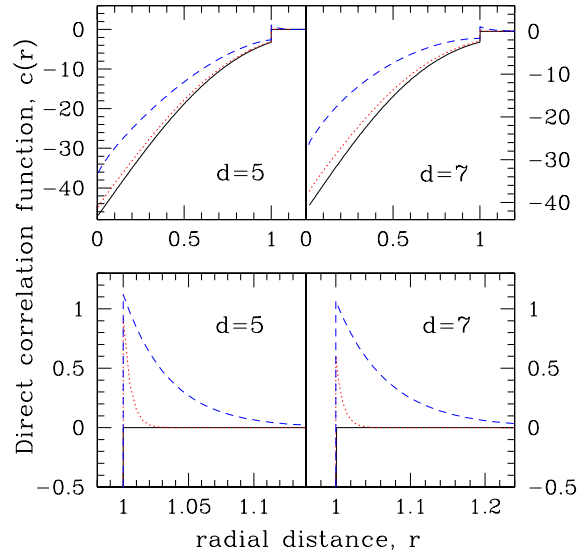}}
\caption{\label{f.yukawa} (Color online) Top panels: Direct
correlation function for $d=5$ (at $\eta=0.2$) and $d=7$ (at
$\eta=0.07$) as computed from the standard RFA (solid lines) and the
extended RFA using the SMS EOS \cite{song} (dotted lines) and the LM
EOS \cite{luban} (dashed lines). Bottom panels: Details of the
Yukawa tails outside the core as obtained from the extended RFA
solutions.}
\end{figure}
\begin{figure}
{\includegraphics[width=.9\columnwidth]{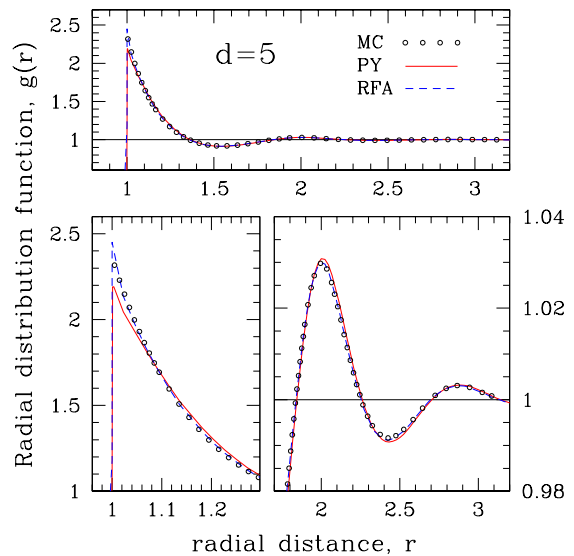}}
\caption{\label{f.gd5_rho08} (Color online) Radial distribution
function of hard hyperspheres in $d=5$ obtained from the extended
RFA method (dashed line), the PY solution (solid line), and Monte
Carlo simulations \cite{bishop} (symbols) at $\rho=0.8$ ($\eta
\approx 0.1316$). The top panel shows the global behavior, while the
bottom panels show the details near the first peak (left) and
between the second and third peaks (right).}
\end{figure}
%
\begin{figure}
{\includegraphics[width=.9\columnwidth]{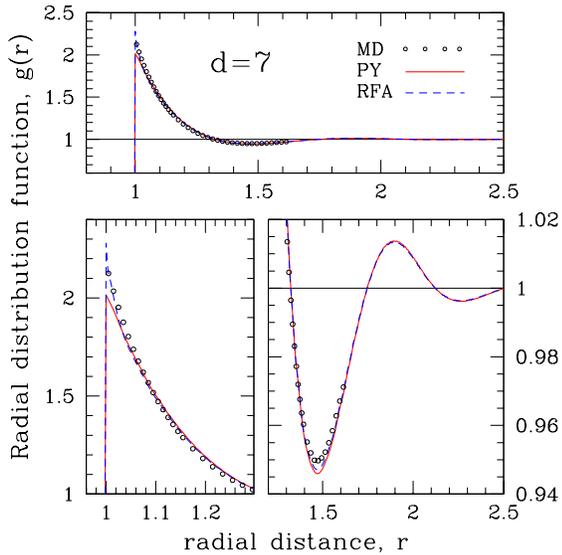}}
\caption{\label{f.gd7_rho14} (Color online) Radial distribution
function of hard hyperspheres in $d=7$ obtained from the extended
RFA method (dashed line), the PY solution (solid line), and
molecular dynamics simulations {\cite{LB06,whitlock}} (symbols), at
$\rho=1.4$ ($\eta \approx 0.05168$). The top panel shows the global
behavior, while the bottom panels show the details near the first
peak (left) and near the first minimum (right).}
\end{figure}
As seen in Sec.\ \ref{s.cr}, the main signature of the extended RFA
solution with respect to the standard one (or PY solution) is that
the former yields a direct correlation function with a (generalized)
Yukawa form outside the core [cf.\ Eq.\ \eqref{cr_ex}]. In Fig.\
\ref{f.yukawa} we compare $c(r)$ evaluated from the standard and
extended RFA (the latter being complemented by the SMS and LM EOS)
for $d=5$ and $d=7$ at densities close to the respective values of
$\eta_f$. At those densities, as shown in Fig.\ \ref{f.z_dvarios},
the SMS EOS is closer to the  PY compressibility route than the LM
EOS. As a consequence, the value of the extended coefficient
$a_{n+2}$ is smaller in the former case than in the latter. This
explains the fact that in Fig.\ \ref{f.yukawa} the curves
corresponding to the extended RFA complemented by the SMS EOS are
much closer to the PY ones (and with a  weaker Yukawa tail) than
those obtained by using the LM EOS. Given the semi-empirical
character of the LM EOS and its excellent agreement with computer
simulations \cite{luban,robles}, it is reasonable to expect that the
true $c(r)$ is better represented by the extended RFA complemented
with the LM EOS than with the SMS EOS. Of course, there are other
EOS proposed in the literature for hard hyperspheres
\cite{maeso,baus,ASV89,BMC99}, but the ones considered in Fig.\
\ref{f.yukawa} are enough for illustrative purposes.

Comparisons between the pair distribution function obtained from
both RFA approaches and available computer simulations
\cite{bishop,LB06,whitlock} are shown in Figs.\ \ref{f.gd5_rho08}
and \ref{f.gd7_rho14} for $d=5$ and $d=7$, respectively.  The
extended RFA results have been obtained with the SMS EOS, but we
have checked that no  significant differences are found if the LM
EOS is used instead. This indicates that $g(r)$ is much less
sensitive  than $c(r)$ to the choice of the input EOS.  Figures
\ref{f.gd5_rho08} and \ref{f.gd7_rho14} show that the standard RFA
(or PY solution) provides an accurate estimate of $g(r)$ for all
radial distances at the considered densities ($\eta\approx
0.7\eta_f$), although some small discrepancies appear near the first
maxima and minima, especially at contact. These deviations are
substantially corrected by the extended RFA.

\section{Concluding remarks} \label{s.concl}

In this work we have generalized the RFA method, originally
developed for three-dimensional hard-sphere fluids
\cite{yuste,YHS96,HYS07}, to hypersphere systems in arbitrary odd
dimensions $d$, providing explicit results for $d\le 11$. This
generalization is not trivial at all. In the application of the RFA
approach, one must  define a functional $G(s)$ of $g(r)$ in Laplace
space,  introduce an auxiliary function $\Psi(s)$ directly related
to $G(s)$, and  approximate $\Psi(s)$ by a rational function (or
Pad\'e approximant), determining the coefficients by the application
of basic consistency conditions.

In the one- and three-dimensional cases the key function $G(s)$ is
the Laplace transform  of $g(r)$ and $rg(r)$, respectively
\cite{yuste_sticky,HYS07}. Thus one might be tempted to define
$G(s)$ for $d=\text{odd}$ simply as the Laplace transform of
$r^{(d-1)/2}g(r)$. Here, however, we have adopted the criterion that
$G(s)$ must be defined as to be closely related to the static
structure factor $S(k)$ of the fluid. This has led us to the
definition \eqref{Gs} and to the relationship  (\ref{S_G}), where
the reverse Bessel polynomials \eqref{theta} play a central role. As
a byproduct, we have derived a general polynomial expression for the
overlap volume of two identical hyperspheres, this quantity
providing the cavity function to first order in density, $y_1(r)$.
Once $G(s)$ has been identified, one needs to define the auxiliary
function $\Psi(s)$ to be approximated by a rational function. Using
the exact knowledge of $G(s)$ to first order in density, Eqs.\
\eqref{Gpsieta}--\eqref{psi1}, it turns out that the natural
definition of $\Psi(s)$ is provided by Eq.\ \eqref{Grfa}. Finally,
$\Psi(s)$ has been approximated by a rational function, Eq.\
\eqref{pade}, the degree difference between the numerator and
denominator being fixed by the exact large-$s$ behavior of $G(s)$.
The coefficients in the Pad\'e approximant for $\Psi(s)$  are
constrained to fit the exact small-$s$ behavior of $G(s)$ or,
equivalently, the small-$k$ behavior of $S(k)$. This representation
reproduces the exact expansion in density of $S(k)$ up to  second
order.

We have called standard RFA to the case in which the number of
parameters in the Pad\'e approximant equals the number of
constraints. It turns out that the associated direct correlation
function vanishes outside the core. Therefore, quite interestingly,
the standard RFA coincides with the solution of the PY closure to
the OZ relation for hard-particle fluids in all odd dimensions
\cite{freasier,leutheusser}. This equivalence between two completely
independent paths allows one to view the PY solution for hard
hyperspheres as the simplest one of a broad class of approximations.
In fact, a more flexible approximation is obtained by adding a pair
of new terms (one in the numerator and the other one in the
denominator) in the Pad\'e approximant for $\Psi(s)$, resulting in
what we have called the extended RFA. Apart from satisfying the
small-$k$ behavior of $S(k)$, the parameters are determined by
requiring thermodynamic consistency with a prescribed EOS, which
must satisfy the inequalities \eqref{Zchi} to ensure that the
extended RFA solution is positive definite.

Comparison with available computer simulations for $d=5$
\cite{bishop} and $d=7$ \cite{LB06,whitlock} shows that the radial
distribution function predicted by the standard RFA (or PY solution)
is rather accurate. On the other hand, there exist certain small
discrepancies (especially near contact) that are satisfactorily
corrected by the extended RFA.

The work presented in this paper is aimed at contributing to our
understanding of the structural properties of hard-hypersphere
fluids and to the mathematical intricacies of their
statistical-mechanical description. Moreover, this work paves the
path to the study of other related systems in $d$ dimensions, such
as sticky hard hyperspheres, square-well particles, or
multicomponent hard-hypersphere fluids. Work is now in progress
along these lines and the results will be published elsewhere.

\begin{acknowledgments}

{We thank Santos Bravo Yuste, Mariano L\'opez de Haro, and Salvatore
Torquato  for helpful comments. We are especially grateful to Marvin
Bishop for his suggestions and  for providing us with tables of the
computer simulation data of Refs.\ \cite{bishop}, \cite{LB06}, and
\cite{whitlock}.} One of the authors (R.D.R.) acknowledges the
Carrera del Investigador Cient\'{\i}fico, Consejo de Investigaciones
Cient\'{\i}ficas y T\'{e}cnicas de la Naci\'{o}n (CONICET,
Argentina). This work has been supported by the SeCyT-UNC
(Argentina) through Grant No.\ 162/06, by the Ministerio de
Educaci\'on y Ciencia (Spain) through Grant No.\ FIS2007--60977
(partially financed by FEDER funds), and by the Junta de Extremadura
(Spain) through Grant No.\ GRU07046.

\end{acknowledgments}

\appendix

\section{Fourier transform using reverse Bessel polynomials} \label{s.mathe}
If the function $\xi(\bm{r})=\xi(r)$ is isotropic,  Eqs.\ (\ref{a})
and \eqref{b} become \cite{sneddon}
\beq \label{fk}
\widehat{\xi}(k)= (2\pi)^{d/2} \int_0^\infty \dd r \,r^{d-1} \xi(r)
\frac{J_{d/2-1}(kr)}{(kr)^{d/2-1}} ,
\eeq
\beq \label{fr}
\xi(r)= \frac{1}{(2\pi)^{d/2}} \int_0^\infty \dd k\, k^{d-1}
\widehat{\xi}(k) \frac{J_{d/2-1}(kr)}{(kr)^{d/2-1}} ,
\eeq
respectively. Here $k=|\mathbf{k}|$ is the magnitude of the wave
vector and $J_\nu(x)$ is the Bessel function of the first kind of
order $\nu$. For half-integer order $\nu=n+1/2$, it is usual to
introduce the spherical Bessel function of the first kind $j_n(x)$
given by
\beq \label{bessel}
j_n(x)= \sqrt{\frac {\pi}{2x}} J_{n+1/2}(x).
\eeq
In order to rewrite the Fourier transform  in a more appropriate
form, we express the spherical Bessel functions as follows:
\beqn \label{j_theta}
j_n(x)&=& \frac{\theta_n(-ix)e^{ix}-\theta_n(ix)e^{-ix}}{2ix^{n+1}}
\cr
 &=& \frac {1}{x^{n+1}} \Im [\theta_n(-ix)e^{ix}] ,
\eeqn
where  the  {reverse Bessel polynomial} $\theta_n(t)$ is defined by
Eq.\ \eqref{theta} \cite{note2}.

A useful integral identity for $\theta_n(t)$, which we have not
found in the literature, is
\beq \label{rec2a}
\theta_n(t)e^{-t}=\int_t^\infty \dd z\,z \theta_{n-1}(z)e^{-z} .
\eeq
By taking the derivative of both sides of Eq.\ (\ref{rec2a}) one
obtains
\beq \label{theo2_1}
\frac{\dd}{\dd t}[\theta_n(t)e^{-t}]=- t \theta_{n-1}(t)  e^{-t}.
\eeq
Next, using Eq.\ (\ref{j_theta}), it is easy  to prove that the
recurrence relation of the spherical Bessel functions \cite{abra},
\beq \label{jrec2}
\frac{\dd}{\dd x} j_n(x) = -\frac{n+1}x j_n(x) + j_{n-1}(x),
\eeq
is recovered.  In passing, from Eq.\ (\ref{theo2_1}) we may note the
recurrence formula
\beq \label{rec3}
\frac{\dd}{\dd t}\theta_n(t)=\theta_n(t) - t \theta_{n-1}(t).
\eeq
Another  recurrence relation  is \cite{carlitz}
\beq \label{rec1}
\theta_n(t)=(2n-1)\theta_{n-1}(t)+t^2 \theta_{n-2}(t).
\eeq
Besides, with the help of Eq.\ (\ref{j_theta}) and expression
(10.1.2) of Ref.\ \cite{abra}, one can find the asymptotic relation
\beq \label{limit}
\frac{i}{2}\left[ \theta_n(ix) e^{-ix} -\theta_n(-ix) e^{ix} \right]
 = \frac{x^{2n+1}}{(2n+1)!!}\left[ 1 + O(x^2) \right],
\eeq
which is used in the evaluation of $y_1(r)$ in Sec.\ \ref{s.vol}.

We return now to the problem of expressing Fourier transforms in odd
dimensions. The functions  $\xi(r)$ and $\widehat{\xi}(k)$ can be
extended to negative $r$ and $k$ as $\xi(-r)=\xi(r)$ and
$\widehat{\xi}(-k)=\widehat{\xi}(k)$, respectively. Then, with the
help of (\ref{bessel}) and (\ref{j_theta}) we can rewrite (\ref{fk})
and (\ref{fr}) as
\beq \label{fourk}
\widehat{\xi}(k)= \frac{(2\pi)^{(d-1)/2}}{k^{d-2}}  i
\int_{-\infty}^{\infty}\dd r\, r \xi(r)\theta_n(ikr) e^{-ikr} ,
\eeq
\beq \label{fourr}
\xi(r) = \frac{(2\pi)^{-(d+1)/2}}{r^{d-2}} i \int_{-\infty}^{\infty}
\dd k\,k \widehat{\xi}(k)\theta_n(ikr)e^{-ikr} ,
\eeq
respectively, where $n$ is defined by Eq.\ \eqref{n}. Finally,
introducing the functional \eqref{Fs}, one arrives at Eqs.\
\eqref{fou_k} and \eqref{fou_r}.

It is worthwhile noting that if $\theta_n(x)$ is replaced by its
polynomial expression (\ref{theta}), then $\cm{F}_n[\xi(x)]$ can be
expressed in terms of the Laplace transforms of $x^{j+1}\xi(x)$ with
$0\le j\le n$,
\beq \label{FL}
\mathcal{F}_n[\xi(x)](s) =\sum_{j=0}^n \omega_{n,j}
s^j\mathcal{L}[x^{j+1}\xi(x)](s),
\eeq
\beq \label{Laplace}
\mathcal{L}[\xi(x)](s) \equiv \int_0^\infty \dd x\,\xi(x) e^{-sx}.
\eeq

\section{Evaluation of $P_{n+1}(r)$}
\label{s.overlap}

In this Appendix we obtain an explicit expression for the
 polynomial $P_{n+1}(r)$ related by Eq.\ (\ref{y1}) to the
first-order cavity function $y_1(r)$. The derivation of $P_{n+1}(r)$
from relations (\ref{residue2}) and (\ref{residue}) is
straightforward but rather tedious. We found it more practical to
use a known expression for the scaled overlap volume in terms of the
normalized incomplete beta function \cite{baus},
\beq \label{Omega}
\Omega_d(r)\equiv \frac{y_1(r)}{y_1(0)}=
\frac{B_{1-r^2/4}((d+1)/2,1/2)} {B((d+1)/2,1/2)}\Theta(2-r),
\eeq
where
\beq \label{Beta_i}
B_{x}(a,b)=\int_0^x \dd t\,t^{a-1}(1-t)^{b-1}
\eeq
 is the incomplete beta function and $B(a,b)=B_{x=1}(a,b)=\Gamma(a)\Gamma(b)/\Gamma(a+b)$ is the beta function.
In the present case, for $d$ odd,
\beq \label{Beta}
B((d+1)/2,1/2)=\frac{2^{(d+1)/2}((d-1)/2)!}{d!!}.
\eeq
{}From (\ref{Omega})--(\ref{Beta}), it is straightforward to obtain
\beq \label{Omega_der}
\frac{\dd\Omega_d(r)}{\dd r} =
-\frac{d!!}{2^{(d+1)/2}((d-1)/2)!}\left(1-\frac{r^2}{4}
\right)^{(d-1)/2}\Theta(2-r).
\eeq
Expanding the binomial and integrating over $r$, one has
\beq
\Omega_d(r)=1-\frac{(2n+3)!!r}{2^{n+2}}\sum_{j=0}^{n+1}\frac{(-r^2/4)^j}{
(2j+1)j!(n+1-j)!},
\label{Omega1}
\eeq
where  it has been implicitly assumed that  $r\leq 2$.

On the other hand, inserting Eq.\ \eqref{y1} into the definition of
$\Omega_d(r)$ we get
\beq \label{Omega2}
\Omega_d(r)=\frac{(2n+3)!!}{2}(2-r)^{n+2}P_{n+1}(r),
\eeq
where we have taken into account that, since $y_1(0)=2^d v_d$, then
$P_{n+1}(0)=2^{-(n+1)}/(2n+3)!!$. {}From Eqs.\ \eqref{Omega1} and
\eqref{Omega2} one finally gets
\beq \label{Pr}
P_{n+1}(r)=\sum_{j=0}^{n+1} p_{n+1,j}r^j,
\eeq
with the coefficients
\beqn \label{P_coef}
p_{m,j}&=& \frac{2^{-(m+j)}}{m!} \left[ \frac{(m+j)!}{j!(2m+1)!!}
  \right. \nn && \left.
-2^{-m} \sum_{l=0}^{[(j-1)/2]} \frac{(-1)^l
(m-1+j-2l)!}{(2l+1)(j-1-2l)!l!(m-l)!}\right],\nn&&
\eeqn
where $[(j-1)/2]$ represents the integer part of $(j-1)/2$.

Equation \eqref{Omega1}  coincides with the expression (3-24) of
Ref.\ \cite{TS06} evaluated in the case $d=\text{odd}$. The novel
contribution of the procedure outlined in Sec.\ \ref{s.vol} is to
show the factorization of the overlap volume [Eqs.\ (\ref{y1}) and
(\ref{Omega2})] into the product of $(2-r)^{n+2}$ and $P_{n+1}(r)$,
which is not evident from Eqs.\ \eqref{Omega}, (\ref{Omega_der}), or
\eqref{Omega1}.

\section{Evaluation of $G_1(s)$\label{app_G1}}
In this Appendix  the function $G_1(s)$ defined by Eq.\
\eqref{G1_y1} is evaluated. We start from the identity
\beq
\int\dd r\, r^m (sr)^j e^{-sr}=-\frac{e^{-sr}}{s^{m+1}}
{(m+j)!}\sum_{l=0}^{m+j}\frac{(sr)^l}{l!}.
\eeq
Consequently,
\beqa
\int\dd r\, r^m \theta_n(sr)
e^{-sr}&=&-\frac{e^{-sr}}{s^{m+1}}\sum_{j=0}^n
{(m+j)!}\omega_{n,j}\nn &&\times\sum_{l=0}^{m+j}\frac{(sr)^l}{l!} .
\eeqa
Finally, making use of Eq.\ \eqref{Omega2}, we get
\beq
\frac{2^{2n+3}}{y_1(0)}\int \dd r\, r
y_1(r)\theta_{n}(sr)e^{-sr}=-\frac{e^{-sr}}{s^{2n+5}}Q_{3n+4}(r,s),
\label{int_y1}
\eeq
where we have called
\beqa
Q_{3n+4}(r,s)&\equiv&(2s)^{2n+3}\sum_{j=0}^n
{(j+1)!}\omega_{n,j}\sum_{l=0}^{j+1}\frac{(sr)^l}{l!}\nn &&-
\frac{(2n+3)!!}{2^{n+1}}\sum_{m=0}^{n+1}\frac{(-1)^{m}(4s^2)^{n+1-m}}{(2m+1)m!(n+1-m)!}\nn
&&\times\sum_{j=0}^n
{(2m+2+j)!}\omega_{n,j}\sum_{l=0}^{2m+2+j}\frac{(sr)^l}{l!}.
\eeqa
Equations \eqref{G1_y1} and \eqref{int_y1} allow us to write
\beq
G_1(s)=s^{-(2n+5)}\left[e^{-s}Q_{3n+4}(1,s)-e^{-2s}Q_{3n+4}(2,s)\right].
\label{G1_Q}
\eeq
The function $Q_{3n+4}(2,s)$ can be further simplified. To that end,
note that, according to Eq.\ \eqref{int_y1},
\beq
\cm{F}_n[y_1(r)](s)=\frac{v_d}{s^{2n+5}}\left[Q_{3n+4}(0,s)-e^{-2s}Q_{3n+4}(2,s)\right].
\label{Y1_Q}
\eeq
Using Eq.\ \eqref{fou_k} and comparing with Eq.\ \eqref{y1_kth} one
obtains
\beq
Q_{3n+4}(2,s)=(-1)^n 4^{n+1}(2n+3)!!\theta_{n+1}^2(s),
\label{Q_{3n+4}}
\eeq
\beqa
Q_{3n+4}(0,s)+Q_{3n+4}(0,-s)&=&(-1)^n 2^{2n+3}(2n+3)!!\nn
&&\times\theta_{n+1}(s)\theta_{n+1}(-s).\nn &&
\eeqa

Equation \eqref{G1} is obtained by inserting Eq.\ \eqref{Q_{3n+4}}
into Eq.\ \eqref{G1_Q}, using Eqs.\ \eqref{G0} and \eqref{aa}, and
calling $Q_{3n+4}(s)\equiv Q_{3n+4}(1,s)$.

\section{Constraints on the Pad\'e coefficients}
\label{s.Pade}
In this Appendix we consider the determination of the Pad\'e
coefficients $\{a_j\}$ and $\{b_j\}$ by application of the physical
constraints \eqref{Psi_ssmall} or, equivalently, \eqref{GsHbis}.
Substitution of Eq.\ (\ref{Grfa}) into Eq.\ (\ref{GsH}), with the
term $e^{s}$  expanded in power series, yields
\beq \label{eqppal}
\sum_{j=0}^M \widetilde{a}_j s^{d+j}=\sum_{j=0}^\infty C_j s^j+
\sum_{j=2}^\infty \left(\sum_{l=0}^{j-2} \gamma_{j-2-l}
C_l\right)s^j,
\eeq
where  $C_j$ is the $j$th coefficient in the series expansion of
$B_N(s)e^s-\an\eta A_M(s)$, i.e.,
\beq \label{Cdef1}
C_j=\sum_{l=0}^j \frac{b_l}{(j-l)!}-\an\eta a_j, \quad 0\le j \le M,
\eeq
\beq \label{Cdef2}
C_j=\sum_{l=0}^{\min(j,N)} \frac{b_l}{(l-k)!}, \quad j \geq M+1,
\eeq
and $\widetilde{a}_j$ and $\gamma_j$ are defined in Eq.\
\eqref{agamma}. {}From the analysis of  Eq.\ (\ref{eqppal}) one
obtains
\beq \label{Ccoef}
C_j=0, \quad 0\le j\le d-1,
\eeq
\beq \label{Ccoefd}
C_j=\widetilde{a}_{j-d}, \quad d\le j\le d+1,
\eeq
\beq \label{Ccoefdm}
C_{d+m} +\sum_{j=0}^{m-2}\gamma_{m-2-j} C_{d+j} =
\begin{cases}
\widetilde{a}_m, &  2\le m \le M,\\
0, &        m \geq M+1.   \end{cases}
\eeq
Taking into account that, as a consequence of Eq.\ (\ref{Ajzero}),
the first $n$ factors $\gamma_j$ with $j=\text{odd}$ vanish, it is
obvious that (\ref{Ccoef})--(\ref{Ccoefdm}) provide a total of
${(3d+1)}/{2}=3n+5$  equations without unknown $\gamma_j$ factors.
Therefore, in agreement with Eq.\ \eqref{Psi_ssmall}, this is the
number of constraints on the  Pad\'e coefficients $\{b_j\}$ and
$\{a_j\}$.

In particular,  Eqs.\ (\ref{Ccoef}) and (\ref{Ccoefd}), together
with the definitions \eqref{Cdef1} and \eqref{Cdef2}, allow one to
express the coefficients $\{b_j\}$ in terms of $\{a_j\}$ for the
cases $N=d$ and $N=d+1$,
\beq  \label{b_new}
b_j= \delta_{j,d}+\an\eta
\sum_{l=0}^{\min(j,M)}\frac{(-1)^{j-l}}{(j-l)!}a_l, \quad 0\le j\le
d,
\eeq
\beq  \label{b_d1}
b_{d+1}=\widetilde{a}_1-1 +\an\eta
\sum_{l=0}^{M}\frac{(-1)^{j-l}}{(j-l)!}a_l,
\eeq
where in Eq.\ \eqref{b_new}  we have taken into account that
$\widetilde{a}_0=1$. Notice that if $N=d$, then $b_{d+1}=0$ and Eq.\
(\ref{b_d1}) provides an expression for one unknown $a_j$ in terms
of the other ones. Expression (\ref{b_new}) can be verified by
induction using (\ref{Ccoef}) and the following algebraic identity,
\beq
\sum_{j=0}^{m-1}\frac{(-1)^j}{j!(m-j)!} = \frac{(-1)^{m-1}}{m!}.
\eeq
%

\section{Solutions of the standard RFA for $d=1$, $3$, and $5$}
\label{s.solutions}

The explicit solutions of Eq.\ (\ref{ec1}) are listed here for hard
$d$-sphere systems in $d=1$, $3$, and $5$ within  the standard RFA
approach presented in Sec.\ \ref{s.RFA}.

\subsection{Hard rods} \label{s.d1}

The one-dimensional case ($d=1$) corresponds to $n=-1$. The
recursive relation (\ref{rec1}) yields $\theta_{-1}(t)=1/t$, so
that, according to Eq.\ (\ref{Gs}),
\beq \label{L_d1}
G(s)=s^{-1} \int_0^\infty \dd r\,g(r)e^{-sr}.
\eeq
Therefore, $sG(s)$ is the Laplace transform of $g(r)$. Taking $n=-1$
in Eq.\ \eqref{Grfa_1} with $\an=1$ [cf.\ Eq.\ \eqref{aa}], one
simply gets
\beq \label{G_d1}
sG(s) =\frac{1} { [\eta + (1-\eta) s] e^{ s} -\eta} .
\eeq
The Laplace transform $sG(s)$ can be easily inverted to obtain the
well-known radial distribution function for hard rods
\cite{frenckel}. Thus, the standard RFA becomes exact for $d=1$.

\subsection{Hard spheres} \label{s.d3}

In the case $d=3$ or $n=0$, Eq.\ \eqref{Gs} becomes
\beq
G(s)=\int_0^\infty \dd r\,rg(r)e^{-sr},
\label{G_s_3D}
\eeq
and so $G(s)$ represents the Laplace transform of $rg(r)$. According
to Eq.\ \eqref{Grfa_1},
\beq \label{G_d3}
G(s) =\frac{e^{-s}}{s^2}\frac{1+a_1 s } { 1
 -12\eta [\phi_3(s)+a_1 \phi_{2}(s) ]} ,
\eeq
where we have taken into account that $\an=-12$ and $a_0=1$.
Equation \eqref{ec1} reduces to $D_1=\widetilde{a}_1=a_1$, whose
solution is
\beq
a_1=\frac{1+\eta/2}{1+2\eta } .
\eeq
{}From here one can easily get
\beq \label{gc_d3}
g(1^+)=\frac{1+\eta/2}{(1-\eta)^2}, \quad \chi
=\frac{(1-\eta)^4}{(1+2\eta)^2} .
\eeq
Equations \eqref{G_s_3D}--\eqref{gc_d3} define Wertheim--Thiele's
exact solution of the PY integral equation for hard spheres
\cite{wertheim,thiele}.

\subsection{Hyperspheres in five dimensions}

For a fluid of hyperspheres in $d=5$ ($n=1$) the definition of
$G(s)$ is
\beq
G(s)=\int_0^\infty \dd r\,r(1+rs)g(r)e^{-sr}.
\label{G_s_5D}
\eeq
Since $a_0=3$ and $\an=240$, Eq.\ \eqref{Grfa_1} becomes in this
case
\beq \label{Gd5}
G(s) =\frac{e^{-s}}{s^2} \frac{3+a_1 s+a_2 s^2 } {1 + 240\eta
[3\phi_5(s)+ a_1\phi_4(s)+a_2\phi_3(s)]} .
\eeq
Equation \eqref{ec1} yields $D_1=\widetilde{a}_1$ and $D_3+\gamma_0
D_1=0$, where, according to Eq.\ \eqref{eg0},
$\gamma_0=\widetilde{a}_2-D_2$. The physical solutions are
\beq
a_1=\frac{ 3-3\eta(1+10 a_2)}{1-6 \eta},
\eeq
\beqn
a_2=\frac{ 1+22\eta+78\eta^2+24\eta^3 +(1-\eta)(6\eta-1)\xi}{20\eta \xi^2},
\eeqn
where $\xi\equiv\sqrt{1+18\eta+6\eta^2}$. The associated contact
value and isothermal susceptibility are
\beqn
g(1^+)=\frac{\xi^3-1+33\eta+87\eta^2+6\eta^3}{60\eta(1-\eta)^3},
\label{g1_d5_2}
\eeqn
\beqn
\label{chi_d5}
\chi=
\frac{(1-\eta)^2}{\xi^2}\left[5(1+6\eta+3\eta^2)-2(2+3\eta)\xi\right].
\eeqn
It can be checked that Eq.\ \eqref{Gd5} yields the same structure
factor as that given by the solution of the PY integral equation for
$d=5$ obtained by a different method \cite{freasier,leutheusser}.


\begin{thebibliography}{99}
\bibitem{noteB}
{Note that the diameter of the two intersecting hyperspheres
considered in the evaluation of $y_1(r)$ is twice that of the true
hard hyperspheres of the fluid.}
\bibitem{note1}
Of course, $\widehat{f}(k)$ can be expressed in terms of the Bessel
function $J_{d/2-1}$ using Eqs.\ (\protect\ref{bessel}) and
(\protect\ref{j_theta}).
\bibitem{s.bN1}
If one adopts the normalization condition \protect\eqref{bN} instead
of
 Eq.\ \protect\eqref{a0},  one gets
%
$b_j=[{a_0}/{(d-2)!!}]\delta_{j,d}+\an\eta \sum_{l=0}^{n+1}
{(-1)^{j-l}}a_l/{(j-l)!}$,
%
with $0\le j \le d$. For $j=d$, the above equation yields a linear
relationship among all $a_j$. Instead of Eq.\
(\protect\ref{Grfa_1}), the standard RFA can then be written in the
equivalent form
%
$ G(s)=s^{-2}{e^{- s}} {A_{n+1}(s) }/[ {1 +\an \eta
s^{-1}\sum_{j=0}^{n+1} a_j \phi_{d-j-1}(s)}]$. Notice that in that
case $g(1^+)=a_{n+1}$.
\bibitem{noteYuk}
In fact, in the classical Debye--H\"uckel approximation for
one-component plasmas the effective potential $V(r)$ is the solution
to the Poisson equation $ \nabla^2
V(r)=-\epsilon_0^{-1}q\delta(\mathbf{r})+k_D^2V(r)$, where $k_D$ is
the Debye wavenumber. In Fourier space the solution is $
\widehat{V}(k)=({q}/{\epsilon_0})({k^2+k_D^2})^{-1}$.
 Now, for $d=\text{odd}$
the inverse Fourier transform becomes
$V(r)=[{(2\pi)^{(d-1)/2}}/{2\epsilon_0}]{\theta_n(k_Dr){r^{2-d}}e^{-k_Dr}}$,
where use has been made of Eq.\ \protect\eqref{fourr} and the
residue theorem. It is in this sense in which we refer to
$\theta_n(\kappa r) e^{-\kappa r}/r^{d-2}$ as a (generalized) Yukawa
term.
\bibitem{note2}
The ordinary Bessel polynomials are given by $y_n(t)=t^n
\theta_n(1/t)$ (cf.\ Ref.\ \protect\cite{bochner}).
\end{thebibliography}

\end{document}